\makeatletter
\@ifundefined{@parse@version@dash}{%
	\def\@parse@version#1{\@parse@version@0#1}
	\def\@parse@version@#1/#2/#3#4#5\@nil{%
		\@parse@version@dash#1-#2-#3#4\@nil}
	\def\@parse@version@dash#1-#2-#3#4#5\@nil{%
		\if\relax#2\relax\else#1\fi#2#3#4 }
}{}
\makeatother

\makeatletter
\DeclareFontFamily{U}{tipa}{}
\DeclareFontShape{U}{tipa}{m}{n}{<->tipa10}{}
\newcommand{\arc@char}{{\usefont{U}{tipa}{m}{n}\symbol{62}}}%

\newcommand{\arc}[1]{\mathpalette\arc@arc{#1}}

\newcommand{\arc@arc}[2]{%
	\sbox0{$\m@th#1#2$}%
	\vbox{
		\hbox{\resizebox{\wd0}{\height}{\arc@char}}
		\nointerlineskip
		\box0
	}%
}
\makeatother

\documentclass[%
reprint,
superscriptaddress,
amsmath,amssymb,
aps,
prb,
showkeys
]{revtex4-2}
\usepackage{verbatim}


\usepackage{siunitx}
\usepackage{color}
\usepackage{graphicx}% Include figure files
\usepackage{dcolumn}% Align table columns on decimal point
\usepackage{bm}% bold math
\usepackage{hyperref}% add hypertext capabilities
\def\d{{\textrm d}}

\def\n{{\mathbf{n}}}

\def\n{\mathbf{n}}
\def\k{\boldsymbol{\kappa}}

\newcommand{\beq}{\begin{equation}}
	\newcommand{\eeq}{\end{equation}}
\newcommand{\beqs}{\begin{eqnarray}}
	\newcommand{\eeqs}{\end{eqnarray}}

\usepackage{amsmath}

\newcommand{\bfb}{{\bf b}}

\newcommand{\bfe}{{\bf e}}

\newcommand{\bfn}{{\bf n}}

\newcommand{\bfr}{{\bf r}}

\newcommand{\bft}{{\bf t}}
\newcommand{\bfu}{{\bf u}}

\newcommand{\bfy}{{\bf y}}

\newcommand{\bfA}{{\bf A}}
\newcommand{\bfB}{{\bf B}}

\newcommand{\bfL}{{\bf L}}

\newcommand{\bfN}{{\bf N}}

\newcommand{\bfX}{{\bf X}}

\def\d{{\textrm d} }

\footnotetext{These authors contributed equally to this work.}

\footnotetext{E-mail: jsb56@cam.ac.uk}

\begin{document}
	
	\title{Geometry, mechanics and actuation of intrinsically curved folds}
	
	\author{Fan Feng$^{\dagger,1}$}
	\author{Klaudia Dradrach$^{\dagger,1}$}
	\author{Michał Zmyślony$^1$}
	\author{Morgan Barnes$^1$} 
	\author{John S. Biggins$^{1,*}$ \\ \vspace{5pt}\it{$^1$Department of Engineering, University of Cambridge,}\\ \it{Trumpington St., Cambridge CB2 1PZ, United Kingdom}}
	%\email{jsb56@cam.ac.uk}

	%\affiliation{Department of Engineering, University of Cambridge, Trumpington St., Cambridge CB2 1PZ, United Kingdom}

	\date{\today}% It is always \today, today,
	%  but any date may be explicitly specified
	
	\begin{abstract}
		We combine theory and experiments to explore the kinematics and actuation of intrinsically curved folds (ICFs) in otherwise developable shells. Unlike origami folds, ICFs are not bending isometries of flat sheets,  but arise via non-isometric processes (growth/moulding) or by joining sheets along curved boundaries. Experimentally, we implement both, first making joined ICFs from paper, then fabricating flat liquid crystal elastomer (LCE)  sheets that morph into ICFs upon heating/swelling via programmed metric changes. Theoretically, an ICF's intrinsic geometry is defined by the geodesic curvatures on either side, $\kappa_{g_i}$. Given these, and a target 3D fold-line, one can construct the entire surface isometrically, and  compute the bending energy. This construction shows  ICFs are bending mechanisms, with a continuous family of isometries trading fold angle against fold-line curvature.  In ICFs with symmetric $\kappa_{g_i}$, straightening the fold-line culminates in a  fully-folded flat state that is deployable but weak, while asymmetric ICFs ultimately lock with a mechanically strong finite-angle. When unloaded, freely-hinged ICFs   simply adopt the (thickness $t$ independent) isometry that minimizes the bend energy. In contrast, in LCE ICFs  a competition between flank and ridge selects a ridge curvature that, unusually, scales as $t^{-1/7}$.  Finally, we demonstrate how multiple ICFs can be combined in one LCE sheet, to create a versatile stretch-strong gripper that lifts  $\sim$40x its own weight.
	\end{abstract}
	

	\maketitle
	
	{\bf Significance Statement.} 
	A foundational result in differential geometry is that bending a surface isometrically changes extrinsic curvature but not intrinsic (Gauss) curvature. Thus intrinsically curved shells cannot be flattened, making them strong.  We explore this dichotomy for curved folds through shells, which pervade biology and engineering. Such folds are generically intrinsic, and  categorized by sign and symmetry. Interestingly, intrinsically curved folds (ICFs) can bend isometrically by trading fold angle vs curvature, with  symmetric folds even reaching a fully flat-folded state. We also create liquid crystal elastomer ribbons that actively morph from flat into ICFs, and combine several such folds in a gripper that lifts $\sim$40x its weight. Our observations highlight ICFs morphing potential, enabling robotics and deployable structures.
	
	\begin{figure}[!h]
		\centering
		\includegraphics[width=\linewidth]{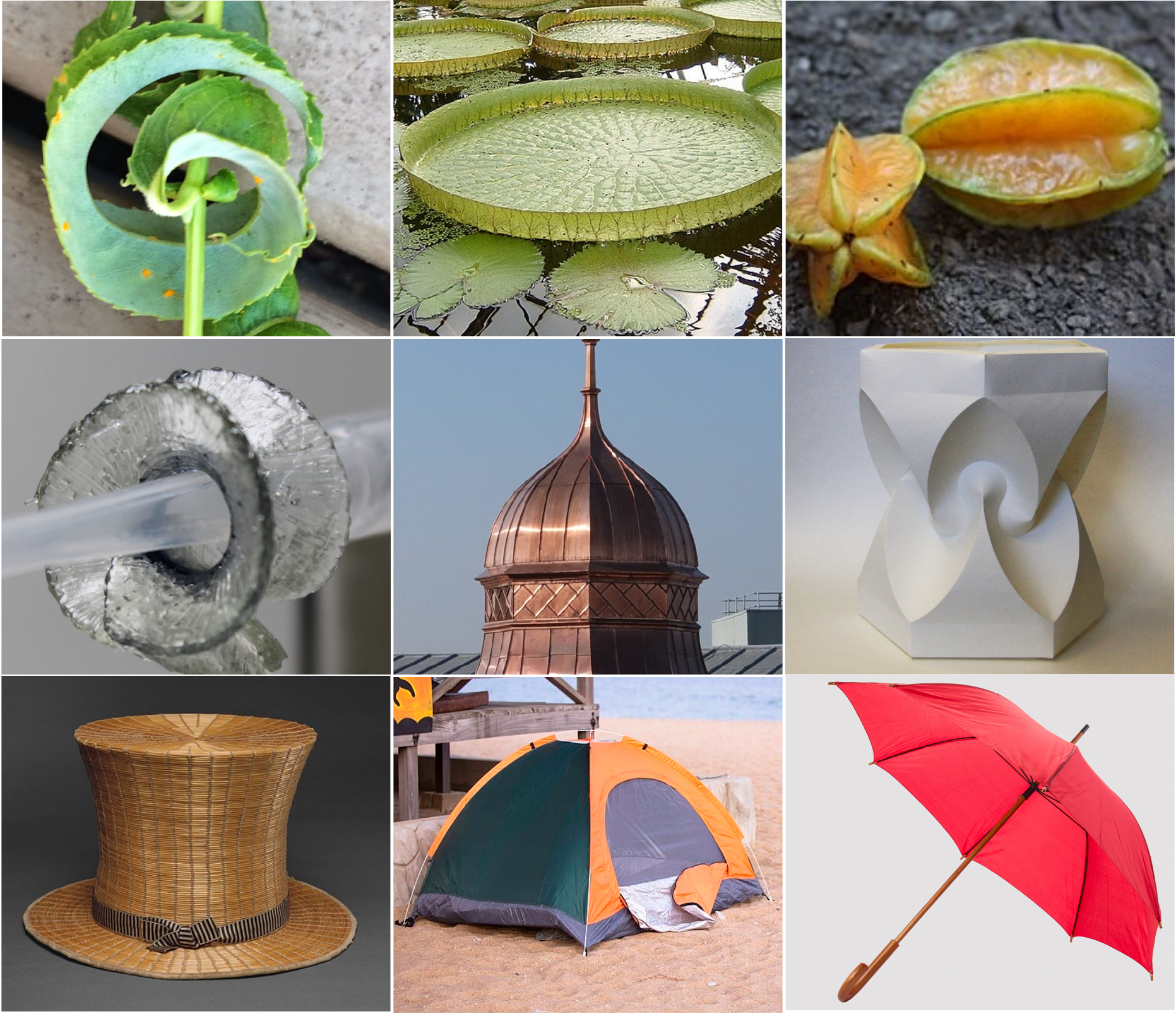}
		\caption{Curved folds in biology, architecture, and engineering. Top: Leaf of \emph{Salix babylonica} ‘Annularis’ (S-) \cite{leaf}, giant water lilly (A+) \cite{lily}, carambola fruit (S+, S-) \cite{star_fruit}.  Middle: LCE ribbon (S-), Cupola, Sedgwick Museum (S+) \cite{roof}, origami tower (D.\ Huffman, \cite{Demaine2016ReconstructingDH, mitani2019curved, origamitower}, extrinsic). Bottom: Top hat (A-) \cite{tophat}, tent (A+) \cite{tent}, umbrella (S+) \cite{umbrella}. Labels (SA/$\pm$) show sign/symmetry of the geodesic curvatures.}
		\label{fig:fig1}
	\end{figure}
	Curved folds lend strength, form and function to surfaces throughout biology and engineering (Fig.\ \ref{fig:fig1}), and have been extensively studied in the beautiful and useful art of origami \cite{duncan1982folded, fuchs1999more, dias2012geometric, mitani2019curved, liu2023design}.  However, not all curved folds are equal. The defining mechanics of thin sheets is that they are essentially inextensible, as isometric bend has a much lower elastic cost than in-plane stretch. Origami artists bend flat paper, making their folds extrinsic (isometric to the plane) but there is no such restriction on folds formed by stitching flat sheets \cite{liu2020tapered, zhai2020situ}, or by non-isometric processes such as differential growth \cite{thomson1917growth, goriely2017mathematics, Paul_non-isometric} or moulding. Such processes can thus create intrinsically curved folds (ICFs), which can only be flattened by stretch, giving richer geometry and stronger mechanics.  Here, we present the basic kinematic rules for describing and classifying ICFs, which provide insight into their (lack of) rigidity, and also into their utility as mechanisms and deployable structures.

	An additional motivation for ICFs stems from the topical field of ``metric mechanics'' \cite{warner2020topographic}, which studies  flat actuating sheets that morph into intrinsically curved surfaces. Such morphing requires a programmed spatial pattern of actuation, reminiscent of  differential growth during morphogenesis, and has been implemented with swelling-gels \cite{klein2007shaping}, phase-changing liquid crystal elastomers (LCEs) \cite{modes2011gaussian,ware2015voxelated, aharoni2018universal}, dielectric elastomers \cite{hajiesmaili2022programmed} and pneumatic baromorphs \cite{siefert2019bio}. The generation of intrinsic (Gauss) curvature gives dramatically strong actuation: for example, LCE disks that morph into  conical shells can lift 1000x their weight \cite{guin2018layered}. Thus inspired, we also demonstrate how to program an LCE sheet to morph into an ICF \cite{duffy2021shape, feng2022interfacial}, leading to flat ribbons that macroscopically ``bend'' into arcs but via strong Gaussian actuation. %The equilibrium shape of a thin sheet is typically the bend-energy minimizing isometry of its metric \cite{klein2007shaping}, which is independent of thickness.  Hinged ICFs follow this rule, but active-sheet ICFs are an exception with equilibrium curvature that diverges in the thin limit.

	\section{Geometry of intrinsically curved folds}
	\begin{figure}%[ht]
		\centering
		\includegraphics[width=\linewidth]{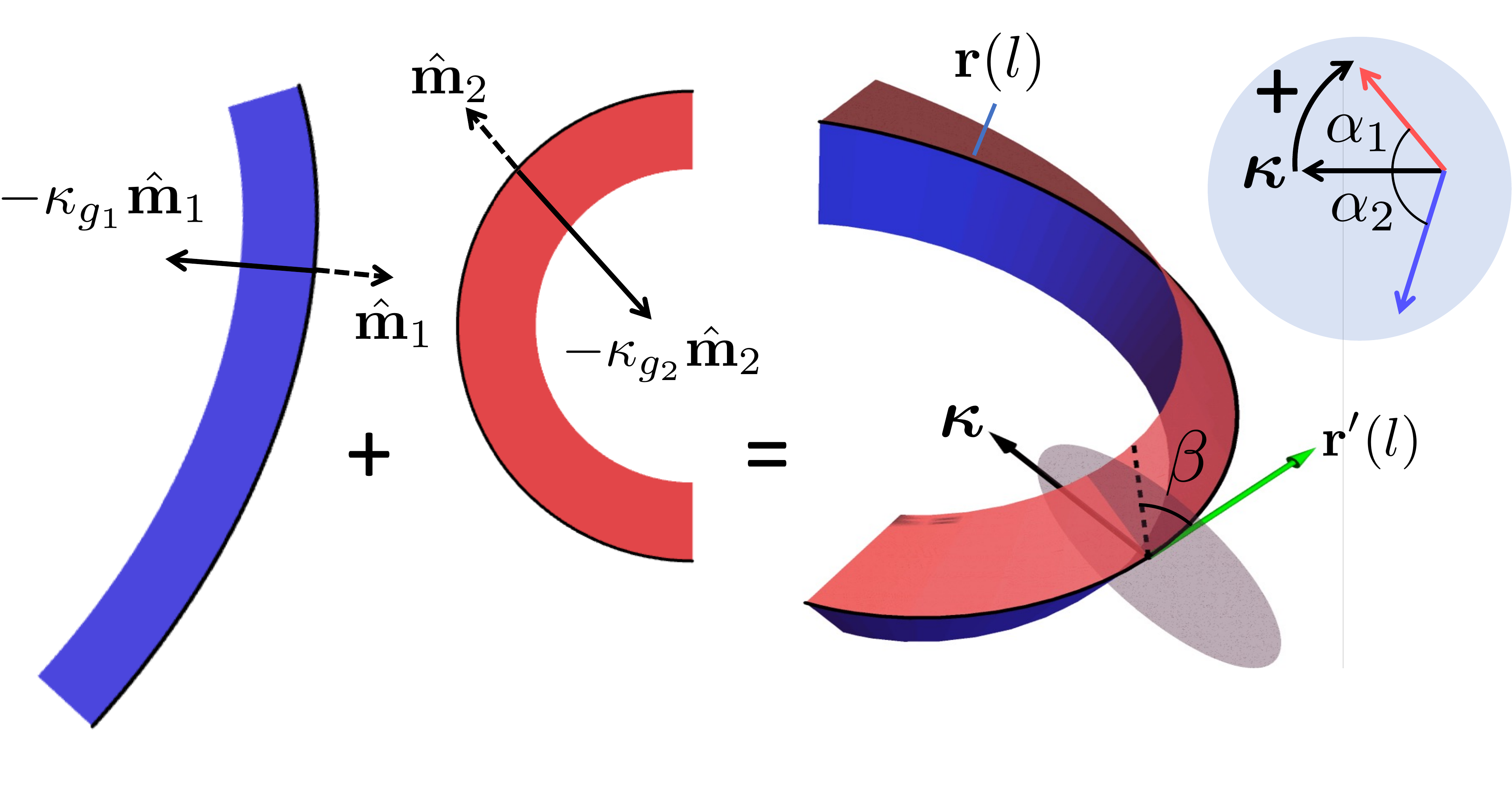}
		\caption{Construction of a general ICF.}
		\label{fig:fig2}
	\end{figure}
	The elementary way to fabricate an ICF is to stitch together two flat sheets of material along curved boundaries. The seam then forms an ICF through an otherwise developable surface 
	(Fig.~\ref{fig:fig2}). This approach is routinely used by engineers, architects and tailors to create intrinsically curved surfaces from flat material. Before stitching, the boundary of each flat piece ($i=1,2$) may be assigned a local outward normal $\mathbf{\hat{m}}_i$ and curvature vector $-\kappa_{g_i} \mathbf{\hat{m}}_i$, so that $\kappa_{g_i}>0$ is an inward curvature vector and vice versa.
	After stitching, both boundaries must follow the same (arc-length parameterized) 3D space curve $\mathbf{r}(l)$ which defines the fold line. The resultant ICF geometry is characterized by the fold's curvature vector $\k=\mathbf{r}''(l)$, and, in a cross-section perpendicular to the fold-line tangent, $\mathbf{r}'(l)$,  the fold angles $\alpha_i$ between $\k$ and each flank. As marked in Fig.~\ref{fig:fig2}, $\alpha>0$ indicates a clockwise rotation from $\k$ to flank (around the tangent), so the total fold angle is $\alpha_1-\alpha_2$.
	
	The geodesic curvature $\kappa_g$ of a curve on a surface is computed as the projection of its 3D curvature vector into the tangent plane. However, a foundational result in differential geometry dictates that $\kappa_g$ is an intrinsic quantity that is invariant under isometric deformations  \cite{o2006elementary}. Accordingly, in the ICF, the geodesic curvatures of the original flat boundaries, $\kappa_{g_i}$, must match the projection of $\k$ into the respective flank:
	\begin{equation}
		\kappa_{g_1} = |\boldsymbol{\kappa}|\cos\alpha_1,\,\,\,\,\,\,\, \kappa_{g_2} = |\boldsymbol{\kappa}| \cos \alpha_2. \label{eq:alpha}
	\end{equation}
	For any proposed fold line with  $|\k|\ge \mathrm{max}\left(|\kappa_{g_1}|,|\kappa_{g_2}|\right),$ one may apply these kinematic relations to compute $\alpha_i(l)$. Furthermore, since the  flanks are developable and hence ruled, their form is then determined, even far from the fold. Precisely, the angle $\beta_i(l)$ between ruling and fold (Fig.~\ref{fig:fig2}), may be computed (\cite{fuchs1999more}, SI Sec.\ 1) in terms of the fold torsion $\tau(l)$: 
	\begin{equation}
		\cot{\beta_i}=\frac{\alpha_i'(l)-\tau(l)}{|\k(l)| \sin\alpha_i(l)}. \label{eq:beta}
	\end{equation}
	
	Another celebrated intrinsic property of surfaces is  Gaussian curvature, $K$, computed as the product of the two principal curvatures. An ICF's Gaussian curvature defies direct computation due to the sharp apex. However, the Gauss-Bonnet theorem allows us to compute the  total curvature $\Omega=
	\int K \mathrm{d}A$ in any patch of surface from the geodesic curvature of its boundary. A simple application to ICFs gives the distribution of total curvature along the fold as \cite{feng2022interfacial}
	\begin{equation}
		\frac{\d \Omega}{\d l} = \kappa_{g_1} + \kappa_{g_2} =|\k|(\cos{\alpha_1}+\cos{\alpha_2}),
	\end{equation}
	so any fold with $ \kappa_{g_1} \ne - \kappa_{g_2} $ is  intrinsic with singular $K$. 
	
	\section{Kinematics and classification of  curved folds}
\begin{figure}%[ht]
		\centering
		\includegraphics[width=\linewidth]{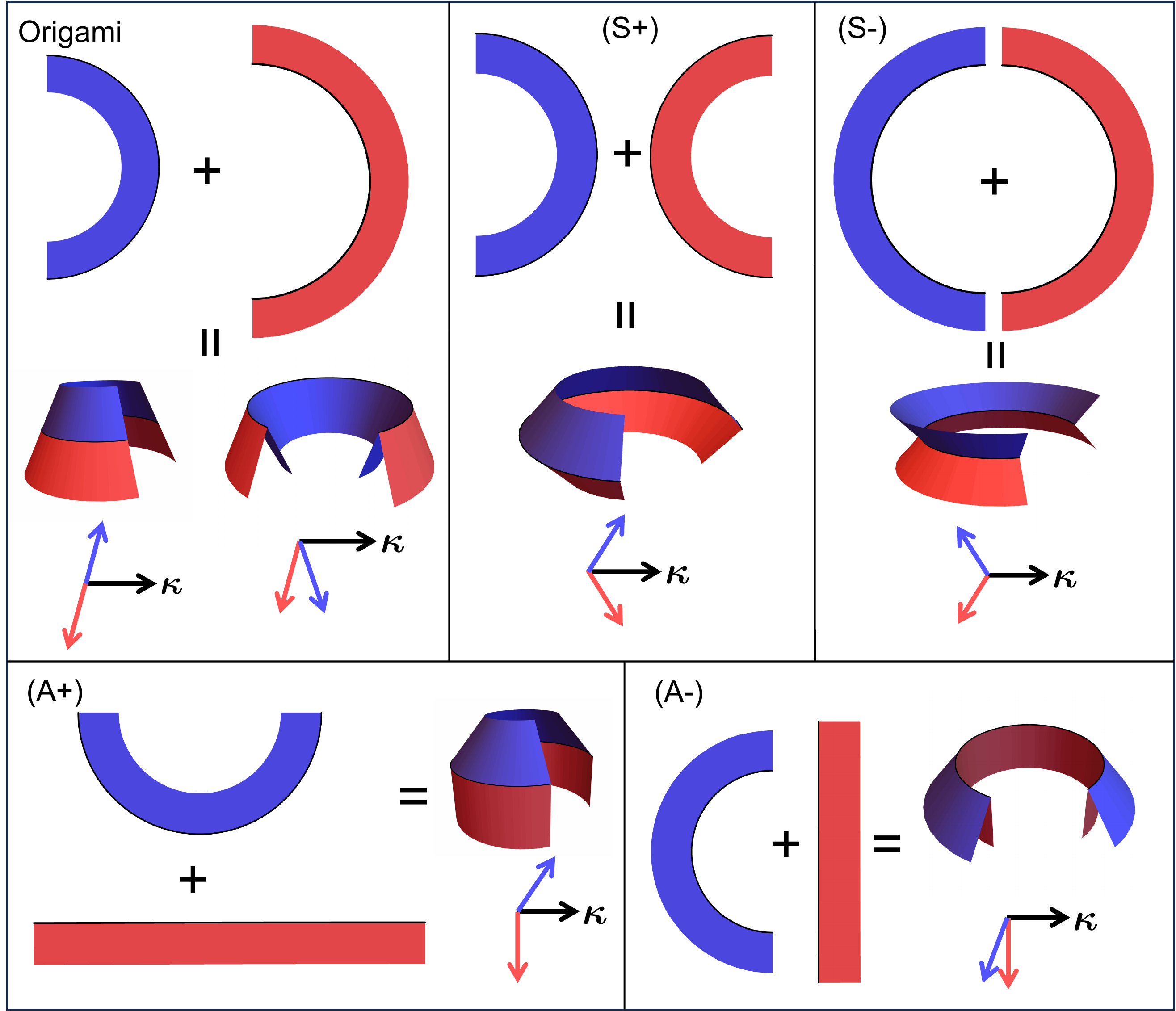}
		\caption{Curved-fold origami and the classification of ICFs.}
		\label{fig:syn_asym}
	\end{figure}

	Eq.\ \ref{eq:alpha} reveals that the kinematics of an ICF are governed by its two geodesic curvatures. To explore what deformation are permitted, we focus on homogeneous ICFs, made by joining pieces with constant $\kappa_{g_i}$ (annular sectors) along the arc of a circle, leading to uniform fold angles. Such ICFs can also be fabricated straightforwardly by joining annular arcs of craft paper (SI Sec. 6), enabling observations of kinematics and mechanics. Five ICF categories emerge, based on combinations of $\kappa_{g_i}$, as illustrated in Fig.\ \ref{fig:syn_asym} and movies M1-M5.
	
	Regular surfaces are often classified by the sign of their Gaussian curvature, with $K=0$ (flat), $K>0$ (cap-like) and $K<0$ (saddle-like) giving different geometry and mechanics.  Accordingly, we first consider a $K=0$ fold (origami case in Fig.\ \ref{fig:syn_asym}) which requires $\kappa_{g_1} =-\kappa_{g_2} \equiv \kappa_g$.  The two initial pieces are thus complimentary, and can fit together whilst flat, confirming the fold is extrinsic. The kinematic constraints (eq.\ \ref{eq:alpha}) give the fundamental rules of origami folds \cite{duncan1982folded, fuchs1999more}, {$\alpha_1=\cos^{-1}(|\k|/\kappa_g)$, and $\alpha_2=\pi- \alpha_1$. The least curved state, $|\k|=\kappa_g$, is the completely flat state ($\alpha_1=0$ and $\alpha_2=\pi$), while, during folding, increasing curvature creates an increasingly sharp fold, with the fold-line  bi-normal bisecting the flanks (movie M1).
		
		Alternatively, an elementary $K>0$ fold can be made by taking $\kappa_{g_1}=\kappa_{g_2}=\kappa_g>0$,  a symmetric positive ICF (S+). Eq.\ (\ref{eq:alpha}) now requires $\alpha_{1}=-\alpha_2=\cos^{-1}(\kappa_g/|\k|)$, so $\k$ itself bisects the flanks, ensuring an equal projection into each (Fig.\ \ref{fig:syn_asym} (S+), movie M2,  Fig.\ \ref{fig:fig1} cupola/umbrella). The least curved state, $|\k|=\kappa_g$, is a flat-folded closed-book configuration, and increasing $|\k|$ requires the book to open, tending towards an unfolded state, $\alpha_1=-\alpha_2=\pi/2$, as $|\k|$ diverges. The fold line and apex curvature have the same sense, as expected for $K>0$ surfaces, and these isometries can be interpreted as trading fold-line curvature and folding angle at constant $K$. 
		%The curvature of the fold line, and the singular curvature of the $V$ have the same sense, qualitatively agreeing with the assignation $K>0$. 
		A corresponding symmetric negative fold (S-) is made as $\kappa_{g_1}=\kappa_{g_2}\equiv-\kappa_g<0$, giving a fold with almost identical behavior (Fig.\ \ref{fig:syn_asym} (S-), movie M3, leaf in Fig.\ \ref{fig:fig1}) except the book-like state is inverted $\alpha_1=-\alpha_2=\pi$ and, as the curvature increases the flanks approach the unfolded $\alpha_1=-\alpha_2=\pi/2$ from above. %In these folds the curvature of the fold line, and the singular curvature of the $V$ have the opposite sense, again agreeing with the assignation $K<0$.
		
		However, ICFs are not fully characterized by their Gaussian curvature, as Eq.\ (\ref{eq:alpha}) applies to each flank individually. Thus folds with asymmetric  curvatures ($A\pm$) behave differently. We focus on two prototypical examples made by joining a straight strip,  $\kappa_{g_1}=0$ with an anulus of each sign $\kappa_{g_2}= \pm \kappa_g$  (Fig.\ \ref{fig:syn_asym} A+ and A-, movies M4, M5). Either way, the folding condition on the straight strip requires $\alpha_1=\pi/2$ so $\k$ has zero projection. The least curved states still have $|\k|=\kappa_g$, giving $\alpha_2=0, -\pi$ respectively, so the $A+$ fold resembles a capped cylinder (Fig.\ \ref{fig:fig1} lilly pad), and $A-$  a flanged pipe (Fig.\ \ref{fig:fig1} top hat). Thus,  asymmetric ICFs have finite fold angles even in their least curved state.  As previously, increasing $|\k|$ causes unfolding, with the highly curved but completely unfolded state $\alpha_1=\pi/2$, $\alpha_2=-\pi/2$ reached as $|\k|$ diverges.

		\begin{figure*}%[ht]
			\centering
			\includegraphics[width=\linewidth]{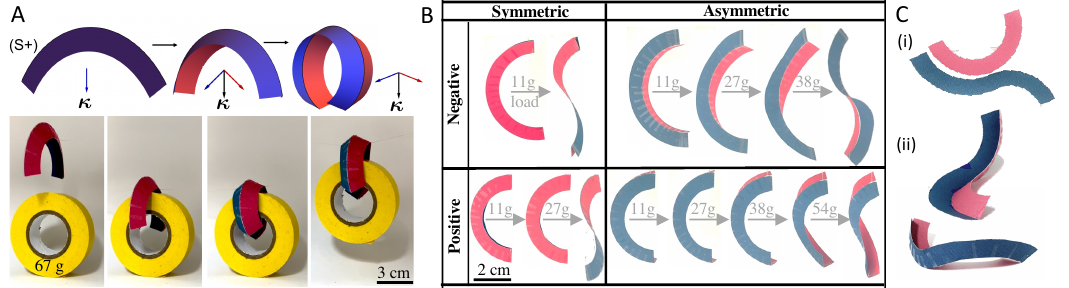}
			\caption{(A) Bending mechanism of the symmetric positive (S+) fold trading curvature vs fold angle, and a corresponding paper grabber. (B) Experimental tensile strength measurements for the four types of folds. (C) (i) Paper strips in 2D with in-homogeneous and asymmetric $\kappa_{gi}$.  (ii) Resulting ICF has a curved ridge with non-zero torsion.}
			\label{fig:kinematics}
		\end{figure*}
		The essential kinematic feature of ICFs is that fold angle dictates curvature and vice-versa, giving a continuous family of isometries in which  the two change in tandem (movies 1-5). ICF's are thus simple mechanisms, and a small actuator controlling fold angle can be used to manipulate the fold's curvature. Furthermore, if the fold-angle is fixed then any deviation of the curvature from its isometric value is strongly resisted. A simple illustration of this  mechanism principle is that a symmetric positive fold fashioned from paper can be used as a simple but effective grabber (Fig. \ref{fig:kinematics}A and movie M6). 
		
		For symmetric folds, unbending of the fold-line (reducing $|\k|$) culminates in a flat-folded state. Symmetric folds are thus attractive deployable structures, which can be constructed  whilst flat, then unfurled into a 3D Gauss-curved surface. 
		
		The different ICF catagories also have different mechanical responses. If a load seeks to unbend an asymmetric ICF,  it will deform isometrically until it locks rigidly in its least curved state --- a 3D shape with finite fold angle --- with further unbending requiring stretch. In contrast, unbending a symmetric fold culminates in the flat-folded state that, mechanically, is a single floppy sheet that can easily buckle to accommodate further unbending. This pattern is confirmed by simple tensile tests (Fig.\ \ref{fig:kinematics}B). Interestingly, both $A+$ and $A-$ folds do eventually buckle out of plane into  inhomogeneous ICF geometries, but $A+$ are considerably stronger than $A-$. The key difference is that further unbending requires tensile stretches in the  annular flank of $A+$, but compression in $A-$, which promote buckling. $A+$ folds are thus particularly suitable for applications requiring rigidity or strong actuation.

		\section{Flank bending energy}
		An ICF formed by joining inextensible flat sheets is limited to the isometric conformations discussed above. However, these configurations are not elastically equivalent, as they have different bends in the flanks. Quantifying  this bend energy allows one to assess which isometry will be observed whilst unloaded, and  the ICF's stiffness as it deforms. When a flank is bent from the initial flat state to form an ICF, it stores elastic energy $E = \int \frac{1}{2} D k^2 \mathrm{d} A$, $D$ being flexural rigidity,  $k$ the single finite curvature of a developable surface, and the integral is over flank area. In the special case of a homogeneous ICF, the whole ICF is a surface of revolution with conical flanks connected by a circular fold. In this case, the bending energy of an individual flank (per-unit length of fold) may be precisely evaluated as
		\begin{equation}
			\rho_{f_i} = \frac{D}{2}  (|\boldsymbol{\kappa}|^2 -\kappa_{g_i}^2) \frac{\log (1- \kappa_{g_i} w_i)}{-\kappa_{g_i}} \approx \frac{D w_i}{2}(|\boldsymbol{\kappa}|^2 -\kappa_{g_i}^2),
		\end{equation}
		where $w_i$ is the width of the flank, and the second form is accurate in the narrow flank limit. As expected, these expressions are zero if $|\boldsymbol{\kappa}|=\kappa_{g_i}$, when the flank lies flat, and penalize curvature in excess of this. If the fold of the ICF is treated as a free-hinge, then the resting configuration will minimize the bending energy of the two flanks, giving $|\boldsymbol{k}|=\mathrm{max}(|\kappa_{g_1}|, |\kappa_{g_2}|)$. In symmetric folds, this corresponds to the fully closed state, while in asymmetric ICFs it gives the locked limiting state. These states are indeed observed as the unloaded equilibria of our paper ICFs.
		
		In general, Eq.~(\ref{eq:beta}) allows one to reconstruct the flanks of any ICF, and then, following Wunderlich \cite{wunderlich1962abwickelbares, todres2015translation, yu2021cutting, starostin2007shape},  compute the associated bend energy as a 1D fold-line integral containing $|\k|$ and $\tau$ (SI Sec.\ 2).   In the narrow limit, this procedure gives
		\begin{align}
			\rho_{f_i} = \frac{D}{2}  w_i \frac{(|\boldsymbol{\kappa}|^2 -\kappa_{g_i}^2 + (\tau - \alpha_i')^2)^2}{|\boldsymbol{\kappa}|^2-\kappa_{g_i}^2},
		\end{align}
		where $\alpha_i$ follows from $|\k|$ via Eq.\ (\ref{eq:alpha}). The unloaded form of the ICF then follows by minimizing the energy of both flanks over $\tau$ and $|\k|$. Interestingly minimizing over $\tau$ gives a simple local condition, which reveals that homogeneous ICFs  ($\alpha_1' = \alpha_2' = 0$) and symmetric ICFs ($w_1=w_2, \kappa_{g_1}=\kappa_{g_2} \implies \alpha_1' = -\alpha_2'$) will form  torsion-free plane curves. However inhomogeneous asymetric ICFs generically do have torsion in their minimizing configuration, generating non-planar fold lines  (Fig.~\ref{fig:kinematics}C).
		
		% \begin{figure}%[ht]
			% \centering
			% \includegraphics[width=\linewidth]{fig_torsion.png}
			% \caption{(A) Paper strips in 2D. (B) Equilibrium configuration with a curved ridge that has non-zero torsion.}
			% \label{fig:torsion}
			% \end{figure}
		\begin{figure*}%[ht]
			\centering
			\includegraphics[width=\linewidth]{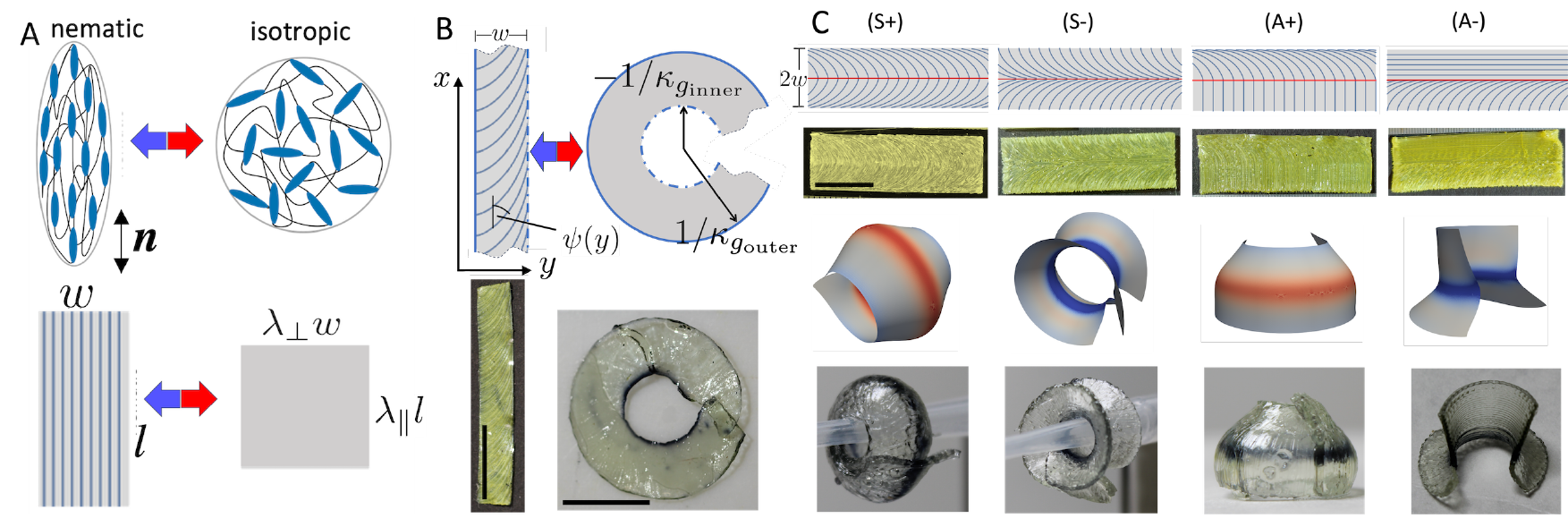}
			\caption{(A) The nematic-isotropic transition in an LCE, which can be induced by heating or swelling with an isotropic solvent. The monodomain nematic director is ${\bf n}$. (B) Top: A planar director pattern in an LCE ribbon that morphs it into an annular arc on actuation. Bottom: 3D printed LCE following the arc-pattern  before and after actuation by swelling in toluene (unactuated thickness $\approx 0.2 \mathrm{mm}$, 2 printed layers). (C) The four types of ICF can be encoded in an LCE by combining  pairs of  arc patterns (S+/-), or the arc pattern with a monodomain (A+/-). From top to bottom: theoretical director patterns, unactuated experimental strips, and actuated ICFs in simulation and experiment. Experiments and simulations  have matched (unactuated) thickness $\approx 0.4 \mathrm{mm}$ (4 printed layers), $w=5\mathrm{mm}$, and actuation factors $\lambda_{\parallel}\approx0.9$, $\lambda_{\perp}\approx2.4$. All scale bars are 10mm.}
			\label{fig:fig_LCEfolds}
		\end{figure*}
		\section{Intrinsically curved folds via metric mechanics}
		ICFs in biological tissues are not formed by stitching of flat sheets, but  by patterns of differential growth that directly alter a single sheet's metric \cite{thomson1917growth}. The emerging field of metric mechanics \cite{klein2007shaping, warner2019topographic} provides an enticing engineering analogue, by programming patterns of shape-change into flat sheets of soft actuating material. ICFs formed via growth or metric-mechanics follow the same kinematics as their joined counterparts, but with the added feature of actuating from flat, and different mechanics stemming from their non-hinged fold. To investigate these differences, we fabricate flat LCE ribbons that morph into ICFs upon stimulation, due to a spatially programmed molecular alignment direction $\n(x,y)\equiv(\cos{\psi}(x,y),\sin{\psi}(x,y))$. The actuation mechanism is that heating or swelling the LCE disrupts this alignment (mirroring the conventional nematic-isotropic phase transition) and causes a large uniaxial shape change, with markedly different stretching factors $\lambda_{\parallel}$ along $\n$, and $\lambda_{\perp}$ in the orthogonal direction, $\n^\perp$ (Fig.\ \ref{fig:fig_LCEfolds}A,B). 
		
		In metric terms, this means an infinitesimal vector  $\mathrm{d}\mathbf{l} = (\mathrm{d}x, \mathrm{d}y)$ in the flat sheet changes length from $\mathrm{d}l^2 = \mathrm{d}l \cdot I \cdot \mathrm{d}l$ to $\mathrm{d}l_A^2 = \mathrm{d}l \cdot \bar{a}(x,y) \cdot \mathrm{d}l$, where the new metric has the form
		\begin{equation}
			\bar{a}(x,y) = \lambda_{\parallel}^2 \n(x,y) \otimes\n(x,y) + \lambda_{\perp}^2  \n^\perp(x,y) \otimes \n^\perp(x,y).
		\end{equation}
		Previous work has considered flat LCEs programmed with pairs of circle \cite{duffy2021shape, feng2020evolving}  or spiral \cite{feng2022interfacial} director patterns, and shown that an in-homogeneous ICF of finite extent appears along the interface between patterns upon actuation. Typically, these ICFs resemble a mountain pass between the conical tips generated by the circle/spiral centers. Here, we instead seek a transitionally invariant pattern to produce an infinitely extendable homogeneous ICF. We thus consider a  flat LCE ribbon that extends infinitely in the $x$ direction, and has a transitionally invariant director profile $\psi(y)$ (Fig.\ \ref{fig:fig_LCEfolds}B). Following our treatment of joined ICFs, we first seek a profile that morphs the ribbon into an annular sector. Since such a sector remains Gaussian flat during actuation, we may apply the \emph{Theorema Egregium} and set $K=0$ to find the profile
		$\psi_c(y) =\frac{1}{2} \arccos(2 y/w - 1)$,\cite{mostajeran2015curvature},
		where $w$ is the width of the ribbon, and the profile varies from along $\hat{\mathbf{x}}$ to $\mathbf{\hat{y}}$ over the the ribbon's width. A direct computation (\cite{duffy2020defective}, SI Sec. 4) gives the geodesic curvatures of the strip's boundaries as
		\[
		\kappa_{g_\mathrm{inner}} =\frac{\lambda_{\parallel}^2 -  \lambda_{\perp}^2}{2 w \lambda_{\parallel}^2 \lambda_{\perp}}, \,\,\,  \kappa_{g_\mathrm{outer}}= \frac{-\lambda_{\parallel}^2 +  \lambda_{\perp}^2}{2 w \lambda_{\parallel} \lambda_{\perp}^2},
		\]
		confirming they follow circular arcs, and hence that the strip becomes annular. Here $\kappa_{g_\mathrm{outer}}>0$ and $\kappa_{g_\mathrm{innter}}<0$ follow the sign convention in Fig.~\ref{fig:fig2}. As sketched in Fig.\ \ref{fig:fig_LCEfolds}C, we may then form positive and negative symmetric ICFs by combining pairs of patterns in a single sheet, with the join becoming the ICF. Similarly, asymmetric ICFs can be created by combining a single pattern with a ribbon of uniform director.
		
		To verify these designs, we fabricated   LCE ribbons via extrusion-based 3D printing  (\cite{saed2019molecularly}, materials and methods), using the extrusion direction to encode the  spatial alignment pattern, and each printed layer  adding  $\approx 100 \mu\mathrm{m}$ of thickness. After printing, actuation was tested on mono-domain ribbons. Swelling  in toluene produced actuation factors of $\lambda_{\parallel}\approx 0.9$ and $\lambda_{\perp}\approx 2.4$ (Fig. S10), while thermal actuation yields $\lambda_{\parallel}\approx0.5$ and $\lambda_{\perp}\approx 1.3$ by 130$^{\circ}\mathrm{C}$ (Fig. S9). Swelling of a ribbon printed with a single copy of the pattern $\psi_c(y)$ indeed produces an annular arc, and ribbons with the four pair-wise combinations of patterns indeed produce the four categories of ICF (Fig.\ \ref{fig:fig_LCEfolds}B,C). The actuated shapes of the ICFs were also computed numerically using the bespoke active-shell C++ code \emph{Morphoshell} \cite{duffy2020defective}, producing excellent agreement with the experiments.
		
		% A direct application of the theoma egrigium \cite{aharoni2014geometry, mostajeran2015curvature, griniasty2019curved} shows that the activated Guass curvature will be
		% \begin{equation}
			% K_A=\frac{1}{2} \left( \lambda_{\perp} ^2-\lambda_{\parallel}^{-2}\right)\left[\sin{2 \psi}\, \partial_y^2 \psi+2\cos{2 \psi} \left(\partial_y \psi\right)^2\right]. 
			% %\half \left(\lambdaperp^2-\lambdapar^{-2}\right)\left[(\partial_y^2 \psi-\partial _x^2 \psi- 4 \partial _x \psi \partial_y \psi)\sin{2 \psi} \right +2 \left(\partial_x\partial_y \psi+(\partial_y \psi)^2-(\partial_x \psi)^2\right)\cos{2 \psi}\right]. \label{CyrusGauss}
			% \end{equation}
		% Solving $K_A=0$ for a Gauss flat annular sector requires
		% \begin{equation}
			% \psi =\frac{1}{2} \arccos(c_1 y + c_2) \\
			% \end{equation}
		% where $c_1$ and $c_2$ are constants of integration. This solution is indeed  only defined in a ribbon of width $w=2/c_1$, and the director transitions from $\hat{x}$ to $\hat{y}$ between the upper and lower edges, which will form the inner and outer arcs of the annular ribbon on activation

		%\textcolor{red}{Experiments and numerics confirm that an individual pattern creates an annular sector, while the paired patterns can create all four categories of ICF.}

		\section{Relaxed shape of metric-mechanic ICFs}
		\begin{figure*}[!ht]
			\centering
			\includegraphics[width=2\columnwidth]{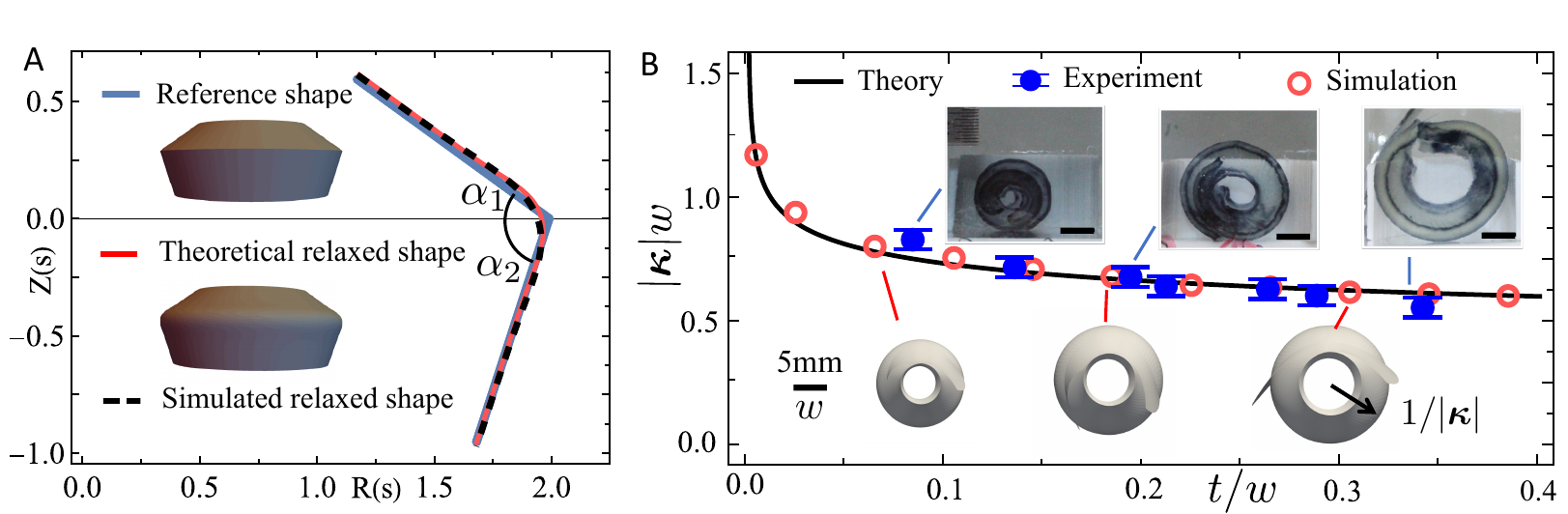}
			\caption{(A) Reference (isometric) shape of an ICF with $\alpha_1=0.2\pi$ and $\alpha_2=0.4 \pi$, and its theoretical (Eq.\ \ref{eq:shape}) and simulated relaxation. (B) Equilibrium ICF curvature against thickness for an (S+) fold: theory (Eq.\ \ref{eq:full_curvature}), experiments, and simulations. All simulation and experiment have equal planar dimensions (reference half-width $w=50\mathrm{mm}$ and length $3 \mathrm{cm}$) and span a range of actuated thicknesses $t$.}
			\label{fig:mechanics}
		\end{figure*} 
		
		Interestingly, the LCE ICFs do not adopt the simple bend-minimizing forms discussed above: for example the symmetric ICFs are far from fully closed. The key difference is that now the fold is  not a freely jointed hinge, so the elastic sheet resists the singular bending deformations required to create a sharp apex. Indeed, one may readily observe in experiments and numerics that the  central fold is not sharp, but rather blunted over some lengthscale $f$, that is short compared to flank-width and radius, but not compared to thickness. Such blunting requires a competition between stretch and bend, with small  strains (deviations from isometry) occurring to relieve the singular curvature of a sharp ridge. To capture this competition, we consider a homogeneous ICF with fixed curvature $|\k|$, as created in our experiments. A perfect isometry of the ICF would form a surface of revolution consisting of two conical flanks connected by a sharp ridge at radius $R_0=1/|\k|$, which we may describe in cylindrical coordinates as $\bar{R}(s)=R_0-|s| \cos{\alpha_i}$, with $s$ being the arc-length transverse to the fold,  $s=0$ being the apex, and $\alpha_i$ switching value between the flanks in accord with Eq.\ \ref{eq:alpha}. Similarly, we describe the blunted form of the ICF by the smooth curve $R(s)=\bar{R}(s)+\Delta R(s)$, and also define the $\theta(s)$ as the angle between the local tangent and the radial direction (see Fig.\ \ref{fig:mechanics}(a)). During blunting, the dominant bend-stretch competition is between $s$ curvature  $\theta'(s)$ (which would diverge at a sharp fold) and hoop strain, $\epsilon=\Delta R/\bar{R}$, leading to the simplified shell energy density
		\begin{equation}
			W=\frac{1}{2} Y \left(\frac{\Delta R}{\bar{R}}\right)^2+ \frac{1}{2}D \theta'(s)^2,
		\end{equation}
		where $Y=3\mu t,~ D=\frac{1}{3}\mu t^3$ are stretching and bending moduli respectively,  $\mu$ being the (incompressible) LCE's shear modulus, and $t$ the actuated thickness. Assuming that the length-scale of blunting is short compared to $R_0$, and also that the isometery is nearly cylindrical so that $\theta'(s)=\Delta{R}''(s)$, we may write the energy of the ridge as 
		\begin{equation}
			E \approx L \int \frac{1}{2} Y \left(\frac{\Delta R}{R_0}\right)^2+ \frac{1}{2} D \Delta R''^2 \mathrm{d} s,\label{eq:energy_cylindrical}
		\end{equation}
		whire $L$ is the length of the ICF. Minimizing  variationally with respect to $\Delta R$ requires $\Delta R^{(4)}(s) + Y \Delta R /D R_0^2$, which admits four independent solutions $\Delta R \propto \exp( (\pm 1 \pm i) s/f)$, revealing $f=(4 R_0^2 D/Y)^{1/4}\sim \sqrt{R_0 t}$ as the emergent blunting length-scale, which matches the blunting lengthscale of (extrinsic) Pogorelov ridges created by mirror inverting a portion of a shell \cite{PogorelovMonograph, VellaPogo, seffeninvertedcones, duffy2023lifting}, and ensures our approximations are self consistent (SI.\ Sec.\ 5). The full form of $\Delta R$ is constructed by taking the decaying solutions on either side of $s=0$, and joining them to produce a smooth and energy minimizing solution, giving
		\begin{align}
			&\Delta R=   -{\textstyle \frac{f}{4}} \mathrm{e}^{-|s/f|} (\cos\alpha_1 + \cos\alpha_2) \left(\cos|s/f| - \sin |s/f| \right)  \label{eq:shape}.
		\end{align}
		Finally, the corresponding vertical position is given by $Z(s)=\int \sqrt{1-R'(s)^2}\mathrm{d}s$. To validate this form, we use Mathematica to numerically minimize a full geometrically-nonlinear energy for an axisymetric shell  (SI.\ Sec. 5) for multiple symmetric ICFs, revealing strong agreement  over a large range of $\alpha$ (Fig.\ \ref{fig:mechanics}, S6).

		Substituting the shape expression into Eq.\ (\ref{eq:energy_cylindrical}),  we obtain the effective ridge energy density $\rho_r=E/L$ as
		\begin{align}
			\rho_r=\frac{1}{4\sqrt{6}} \mu t^{5/2} |\boldsymbol{\kappa}|^{-3/2}(\kappa_{g_1} + \kappa_{g_2})^2.
			\label{eq:ridge_density}
		\end{align}
		Interestingly, this ridge energy again has the same thickness scaling as a Pogorelov ridge \cite{PogorelovMonograph, duffy2023lifting}. However, the ICF energy also includes the signature of the folds Gaussian curvature $(\kappa_{g_1} + \kappa_{g_2})$, and scales with $|\k|^{-3/2}$, showing the ridge   favours less folded more curved configurations. 
		
		To predict the relaxed shape of an ICF, we take the total energy as the ridge energy plus the previously computed bending energy of the two flanks. Minimizing this total energy  $\rho_r + \rho_{f_1} + \rho_{f_2}$ of the fold over curvature yields
		\begin{align}
			& |\boldsymbol{\kappa}| = \label{eq:full_curvature}  \\& \frac{3^\frac{3}{7}}{2} t^{-\frac{1}{7}}(\kappa_{g_1}+\kappa_{g_2})^\frac{4}{7} \left( \frac{\log(1-\kappa_{g_1}w_1)}{\kappa_{g_1}}+ \frac{\log(1-\kappa_{g_2} w_2)}{\kappa_{g_2}} \right)^{-\frac{2}{7}} \notag ,   
		\end{align}
		which, for the narrow symmetric case reduces to:
		\beq
		|\boldsymbol{\kappa}| =(3^3/2^5)^{1/7} \kappa_g^{4/7} t^{-1/7} w^{-2/7}.
		\label{eq:simplified_curvature}
		\eeq
		These results reveal an unusual behavior: the selected isometry is thickness dependent,  with thinner sheets curving more and folding less despite having the same metric. To validate this effect, we numerically (Morphoshell) and experimentally actuate S+ LCE ICFs spanning a magnitude of thickness. Experimentally, thickness was varied via changing the number of printed layers, and actuation was again by swelling in toluene.  As seen in Fig.~\ref{fig:mechanics}, both experiment and numerics clearly exhibit higher curvature at lower thickness, in very satisfactory agreement with Eq.\ \ref{eq:full_curvature}. 
		
		\begin{figure*}[ht]
			\centering
			\includegraphics[width=\linewidth]{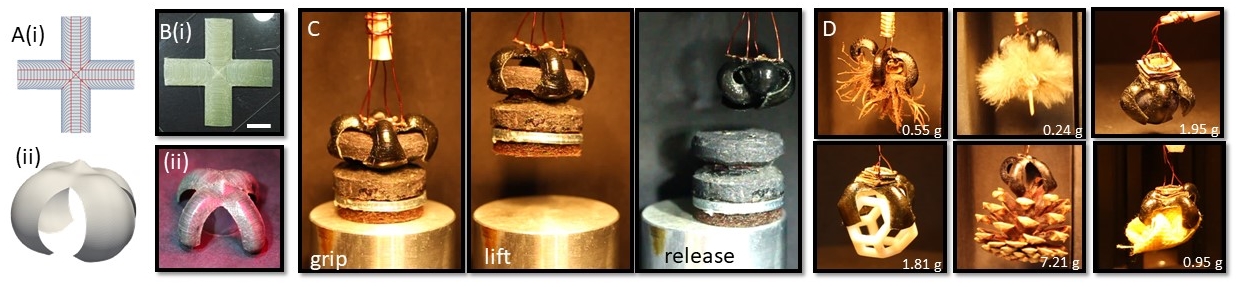}
			\caption{The design of an LCE Gaussian gripper. A(i) Reference director pattern, encoding several ICFs. (ii) Simulated actuated configuration makes a strong claw shape. B: 4-layer 3D printed sample weighing 0.48g (i) before and (ii) after actuation with an IR heat lamp. (C) Controlled grip, lift and release of a weighted cork. Release occurs shortly after the stimulation is turned off. (D) The gripper can lift a wide range of object: clockwise, turkish hazelnut husk, silverpuff, blueberry, fabric, pinecone and 3D printed dodecahedron frame. Scale bar on B(i) is 1 cm.}
			\label{fig:fig_application}
		\end{figure*}
		This behavior contrasts with the general rule for shape selection in thin sheets, in which one expects bend to only enter as a tie breaker between zero-stretch states. Since the bend energy only depends on thickness via a $t^3$ prefactor, the resultant bend-minimising isometry is thickness independent: for example, a given (anti-)conical metric always generates the same (anti-)cone, independent of thickness \cite{guin2018layered}, as would a pattern generating a saddle or spherical cap. Freely hinged ICFs follow this usual rule, as do LCE sheets encoded with either finite Gaussian curvature, or pointwise concentrations at (anti-)conical tips. However, beyond metric mechanics, similar thickness effects are seen in curved-fold origami, if the hinge is formed from an angular spring, setting up a contest between spring energy (thickness independent) and flank bend ($t^3$) to determine which isometry is observed \cite{dias2012geometric, DIAS201457}. Similarly, the thickness dependent behavior of ICFs (line-like curvature concentrations) emerges from the contest between the stretch/bend energy of the blunted fold  ($t^{5/2}$) and the bend energy of the flanks ($t^3$). However, the two effects are fundamentally different: origami creases only fold because of the hinge-spring motivates finite fold angles, leading thinner sheets fold more as the spring becomes more significant. In contrast, ICFs fold because of their intrinsic geometry, with the ridge inhibiting folding so that thinner sheets fold less.

		\section{Discussion}
		Our principle results on the rich kinematics of ICFs highlight how, provided they are constructed from bendable sheets, they form useful mechanisms, with potential applications across soft robotics and morphing/deployable structures. Particular highlights include symmetric ICFs which, pseudo-paradoxically, allow a truly Gauss curved surface to be deployed from a planar flat-packed state, and also asymmetric positive ICFs which can show great strength and rigidity. Our results also highlight the similarities and differences between hinged ICFs and those formed by morphing active sheets, with both obeying the same basic kinematics but having quite different mechanics and shape selection.

		Our treatment has focused on shells containing single ICFs, but a natural extension is to combine multiple ICFs for more complex morphing. As a simple demonstration, we design and fabricate an LCE Gaussian-gripper, by combining multiple strong asymmetric positive folds in a cross-like configuration,  (Fig.~\ref{fig:fig_application}A-B). Upon convenient thermal actuation by a heat lamp,  the cross morphs into a claw shape that is capable of gripping and lifting a simple load (Fig.~\ref{fig:fig_application}C) up to 40x the  gripper's own weight (Movie M7); the gripper is also versatile, lifting a wide range of objects with different shapes, weights, textures and levels of compliance (Fig.~\ref{fig:fig_application}D, Movie M8).

		Looking ahead, our results also motivate many directions for further exploration. Can one design an ICF that traces an arbitrary 3D space curve in its equilibrium state, or even an elephant's-trunk ICF that can morphs between many space-curves via angular actuators along its length? What is the effective rod-theory \cite{DIAS201457, dias2016wunderlich} for ICF ribbons? How do ICF kinematics change when they pass through intrinsically curved ribbons \cite{efrati2015non}?  What are the kinematics of surfaces containing  multiple ICFs, such as our gripper? Overall, the rich geometry and mechanics of ICFs coupled with their straightforward manufacture, suggests these questions, any many others, will be a very fruitful area for further exploration.

		\section{Materials and Methods}
		\subsection{Materials}
		LC ink was synthesised by creating oligomers via aza-Michael addition from a reactive mesogen (RM82; CAS: 125248-71-7), amine linker (n butylamine, nBA; CAS: 109-73-9) and photoinitiator (Ir2959; CAS: 106797-53-9). RM82 was purchased from Daken Chemical, China, and both nBA and Ir2959 from Merck Life Science, UK. All compounds were put to a glass vial (with molar ratio of RM82 to nBA: 1.1:1.0 and with 1 wt $\%$ of initiator), melted with a heat gun and vigorously mixed, first by vortex mixer and then on a hot plate with magnetic stirrer bar for ca. 30 min. A (pre-heated) metal syringe for printing was then filled with the ink and left in the oven for around 18 h for full oligomerisation (\SI{75}{\degreeCelsius} - \SI{80}{\degreeCelsius}). 
		
		\subsection{3D-printing of ICFs}
		The syringe with the ink was mounted in Hyrel’s KR2-15 used with System 30M 3D printer with a 2D array of UV LED (\SI{365}{\nano\metre}). All the patterns were printed on glass slides coated with 5 \% solution of poly(vinyl alcohol) in water, after careful calibration of the printing parameters for \SI{0.30}{\milli\metre} MK8 nozzle (printed line width: \SI{0.20}{\milli\metre}, first layer height: \SI{0.125}{\milli\metre}, layer height: \SI{0.1}{\milli\metre}, printing speed: \SI{260}{\milli\metre\min^{-1}}, extrusion rate: \SI{\sim5.2}{\micro\liter\min^{-1}}, priming: 40 000 - 50 000 pulses with rate 10 000 pulses/s, where 1297 pulses correspond to \SI{1}{\micro\liter}; temperature of printing: \SI{80}{\degreeCelsius}), with UV LEDs working on 30\% - 50\% of max power. After printing all samples were additionally cured for 1 h in a UV curing station with two UV LEDs (LuxiGen LZ1-10UV0R, \SI{365}{\nano\metre}) irradiating a sample placed on a distance of 65 mm from both top and bottom diode. Printed LCE sheets were then detached from their glass substrates with a blade.
		
		\subsection{Actuation}
		Initial thermal strain data was collected on monodomain (linear) samples, using a hot plate (Fig. S9). Swelling strain data was obtained by placing monodomain samples in a  glass container filled with toluene for around 3 h  (Fig. S10). ICF samples were actuated by swelling in the same way, and then imaged  using a Canon EOS 1200D camera, with either EFS 18-55 mm lens or with macro 100 mm lens. ICFs in Fig. \ref{fig:fig_LCEfolds}C were imaged shortly after removal from toluene. The precisely characterized  ICFs in Fig. \ref{fig:mechanics}B were imaged whilst still in toluene, with the camera directly facing the axis of the fold's surface of revolution for accurate extraction of the ICF radius. Thickness, before and after actuation, was measured using digital calipers (Hilka \SI{0}{\milli\metre} - \SI{150}{\milli\metre}). The LCE grippers were actuated thermally with a heat lamp (Panasonic, \SI{300}{\watt}) in a custom box coated internally with 
		0.4 mm thick black paper. The LCE sample was prepared by 3D printing, then painted with a black marker to aid absorption, and the lamp was directed at an angle of \SI{30}{\degree} from the vertical. The same camera was used to record experiments of grasping and lifting different objects with a weight range \SI{0.24}{\gram} - \SI{19.24}{\gram}.  Temperatures up to \SI{150}{\degreeCelsius} were recorded during heat lamp irradiation, using a digital thermometer (CELTEK T2001) with the probe placed between a gripper and a weight.

		\begin{acknowledgments}
			This work was supported by a UKRI ‘future leaders
			fellowship’ grant (grant no. MR/S017186/1). Additionally, M.Z. received funding from the European Union’s Horizon 2020 research and innovation programme under the Marie Skłodowska-Curie grant agreement No 956150.
		\end{acknowledgments}
		\bibliography{curved_fold}% Produces the bibliography via BibTeX.
		%\newpage
		%\detailtexcount{main}
	\end{document}

\title{Supplemental Information: \\Geometry, mechanics and actuation of intrinsically
curved folds}% Force line breaks with \\
%\thanks{A footnote to the article title}%
\author{Fan Feng}
\author{Klaudia Dradrach}
\author{Michał Zmyślony}
\author{Morgan Barnes} 
\author{John S. Biggins}
\affiliation{Department of Engineering, University of Cambridge, Trumpington St., Cambridge CB2 1PZ, United Kingdom}

\date{\today}% It is always \today, today,
             %  but any date may be explicitly specified

\maketitle

%\tableofcontents

%%%%%%%%%% Prefix an "S" to all equations, figures, tables and reset the counter %%%%%%%%%%
\setcounter{equation}{0}
\setcounter{figure}{0}
\setcounter{table}{0}
\setcounter{page}{1}
\setcounter{section}{0}
\makeatletter
\renewcommand{\theequation}{S\arabic{equation}}
\renewcommand{\thefigure}{S\arabic{figure}}
\renewcommand{\thetable}{S\arabic{table}}
\renewcommand{\bibnumfmt}[1]{[S#1]}
\renewcommand{\citenumfont}[1]{S#1}
%%%%%%%%%% Prefix an "S" to all equations, figures, tables and reset the counter %%%%%%%%%%

\section{Isometries of intrinsically curved folds}\label{sec:SI_isometry}
\begin{figure}[h]
	\centering
	\includegraphics[width=\columnwidth]{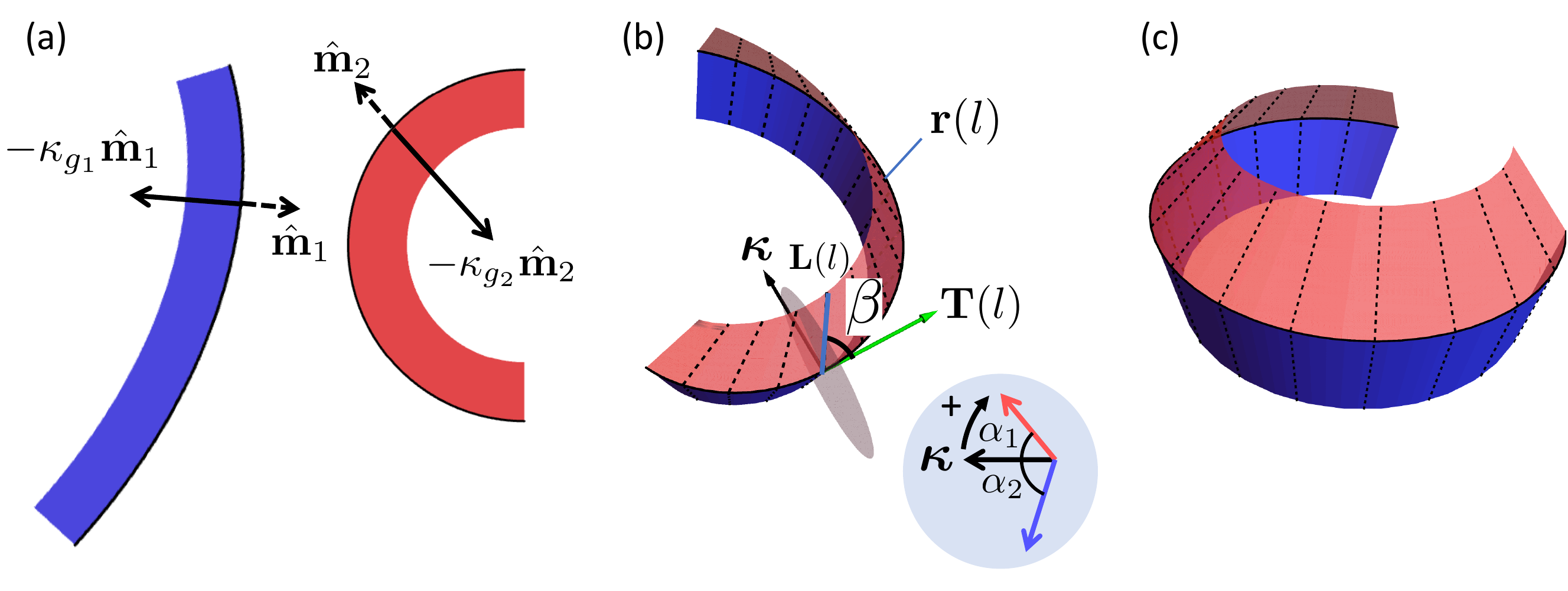}
	\caption{(a) Reference strips in 2D. (b) Isometry of a folded configuration. (c) A different isometry.}
	\label{fig:SI_isometry}
\end{figure}

Inspired by the work on curved-fold origami \cite{duncan1982folded, fuchs1999more}, we  compute the exact form of an ICF by assuming that the deformation is isometric and the flanks are developable surfaces.  More precisely, given the reference strips (Fig.~\ref{fig:SI_isometry}(a)) with geodesic curvatures $\kappa_{g_i}(l)$, the folded configuration Fig.~\ref{fig:SI_isometry}(b) is uniquely determined by the arclength parameterized fold (ridge) $\bfr(l)$. 

To compute the shape explicitly, we first establish two important orthonormal frames. The ridge line itself is associated with a unit tangent vector, $\bft(l)=\bfr'(l)$ and orthogonally the curvature vector $\k=\bfr''(l)$, which together define the curve's osculating plane. We may then define the  Frenet–Serret frame associated with the ridge-line alone, $\{\bft(l),\bfn(l), \bfb(l)\}$, where $\bfn(l)=\k(l)/|\k|$ is the normalized curvature vector (normal vector) of the ridge and $\bfb(l) = \bft(l) \times \bfn(l)$ is the bi-normal. Secondly, we have the Darboux frame $\{\bft(l), \bfN(l), \bfB(l)\}$ associated with the curved ridge $\bfr(l)$ and one flank surface (for example the red surface in Fig.~\ref{fig:SI_isometry}), where $\bft(l)$ is again the unit tangent of the ridge, but $\bfN(l)$ is the unit normal to the flank surface and $\bfB(l) = \bft(l) \times \bfN(l)$ is a vector in the surface orthogonal to the tangent, and hence pointing away from the ridge. In the Darboux frame, $\bfB(l) $ and $\bft(l)$ thus define the tangent plane of the flank surface.

Since  $\bfN$ is orthogonal to $\bft$, we may  express it in the Frenet-Serret frame as $\bfN(l) = \bfb(l)  \cos\alpha  + \bfn(l) \sin\alpha$. The geodesic argument in the main text then gives the fold angles in terms of the curvature:
\begin{equation}
\alpha= (\pm)\arccos(\kappa_{g_i}/|\boldsymbol{\kappa}|).\label{eq:alpha}
\end{equation}

As the flank is developable, it will be a ruled surface, so its form can be  generically expressed as
\beq
\bfX(l, v) = \bfr(l) + v \bfL(l), \quad v \in [0,V_0], \quad l \in [0, L_0]\label{eq:dsurface}
\eeq
where $\bfL(l)$ is the ruling (or generator) of the surface, $V_0(l)$ is the length of the ruling, and $L_0$ is the length of the ridge. Furthermore, since the ruling must lie in the surface, we may express it in the Darboux frame as
\beq
\bfL(l) = \bft(l) \cos\beta(l) + \bfB(l) \sin\beta(l),
\eeq
However, not all ruled surfaces are developable. For the flank to be developable, it must satisfy the additional condition $\bft(l)\times \bfL(l) \cdot \bfL'(l)=0$, which requires the tangent plane to be invariant along the ruling. After substituting the above form into this condition, a direct algebraic computation \cite{yu2021cutting} reveals that the flank is only developable if:
\beq
 \beta(l)=\arccos\left[(\tau-\alpha')/\sqrt{(\tau-\alpha')^2+\kappa^2\sin^2\alpha} \right],
\eeq
or equivalently
\beq
 \beta(l)=\arccot\left[(\tau-\alpha')/(\kappa \sin\alpha) \right], \label{eq:beta}
\eeq
where $\beta(l) \in [0,\pi]$, and $\tau(l)=-\bfn(l) \cdot \bfb'(l)$ is the curve torsion. Having thus computed $\beta(l)$, the shape of the flank is then fully specified by (\ref{eq:dsurface}), and similarly for the shape of the other flank. 

To ensure a solution for a given proposed ridge line, we must have $\cos\alpha \in[-1,1]$, giving the constraint $|\boldsymbol{\kappa}| \geq |\kappa_{g_i}|$ which restricts the configuration space and the isometry of the folded ICF. Figure \ref{fig:SI_isometry}(c) shows a different isometry that satisfies this constraint. Recalling the sign convention for $\alpha_i$ (also shown in Fig.~\ref{fig:SI_isometry}(b)), we have the opening angle given by $\alpha_1-\alpha_2$, which is larger for a more curved ridge with larger $|\boldsymbol{\kappa}|$.

\section{Flank bending energy}
\label{sec:flank} 
 The bending energy per unit area (bending energy density) of a developable surface is proportional to the mean curvature square $H^2$ \cite{Seguin2020ClosedUK, todres2015translation}. Given the explicit form of the developable surface $\bfX(l,v)$, i.e. Eq.~(\ref{eq:dsurface}), the mean curvature is given by \cite{yu2021cutting,wilson_1937} 
 \beq
H = \frac{\sqrt{|\boldsymbol{\kappa}|^2-\kappa_g^2}}{2\sin\beta(\sin\beta+v ( \kappa_g -\beta' ))}.
\eeq
Notice that the generator $\bfL(l)$ in Eq.~(\ref{eq:dsurface}) is a unit vector here for our convenience while the generator direction in \cite{yu2021cutting, starostin2007shape} is not normalized, leading to a slightly different presentation.
 The total bending energy is then given by $E_b = \frac{D}{2} \iint (2H)^2 \d A$, where the area element $\d A$ is computed by $\d A= |\partial_v\bfX \times \partial_l \bfX| \d v \d l =(\sin\beta  + v (\kappa_g - \beta'))\d v \d l$ and $D=E t^3/[12(1-\nu^2)]$ is the bending stiffness of a plate with Youngs modulus $E$, thickness $t$ and Poisson ratio $\nu$.

Integrating the bending energy density along the generator yields the dimension-reduced energy, known as the Wunderlich functional \cite{wunderlich1962abwickelbares}, which is given by \cite{yu2021cutting}
\begin{align}
E_{b}&=\frac{D}{2} \int_0^{L_0} \left(\int_0^{V_0}(2H)^2 (\sin\beta  + v (\kappa_g - \beta'))\d v \right)\text{d} l \nonumber \\
&= \int_0^{L_0} \frac{D}{2} \frac{\kappa^2 - \kappa_g^2}{- (\kappa_g - \beta') \sin^2\beta} \log [1- (\kappa_g - \beta')w/\sin^2\beta ] \text{d} l 
\label{eq:bending_energy} 
\end{align}
where $\kappa=|\boldsymbol{\kappa}|=|\bfr''(l)|$ is the scalar curvature of the ridge. In the second expression, we introduce $w=V_0 \sin\beta$, which, in the narrow limit is the width of the flank,  a   more physically relevant quantity than the generator length $V_0$. For the homogeneous case, we have $\beta=\pi/2$, which reduces this expression to the conical version given in the main text (Main text Eq. 4). In the narrow ribbon limit ($w \kappa_g<<1$), we may take the Taylor expansion of Eq.~(\ref{eq:bending_energy}) and simplify the bending energy as
\beq
E_{\text{narrow}} = \frac{D}{2} \int_0^{L_0} \frac{\kappa^2 - \kappa_g^2}{\sin^4\beta} w \d l = \frac{D}{2}  \int_0^{L_0}  w_i \frac{(|\boldsymbol{\kappa}|^2 -\kappa_{g_i}^2 + (\tau - \alpha_i')^2)^2}{|\boldsymbol{\kappa}|^2-\kappa_{g_i}^2}  \d l,
\eeq
where the second equality follows after substitution for $\beta$ from Eq.\ \ref{eq:beta}, and corresponds to the inhomogeneous narrow result in the main text (Main Text, Eq. 5).

\begin{figure}[h]
	\centering
	\includegraphics[width=\columnwidth]{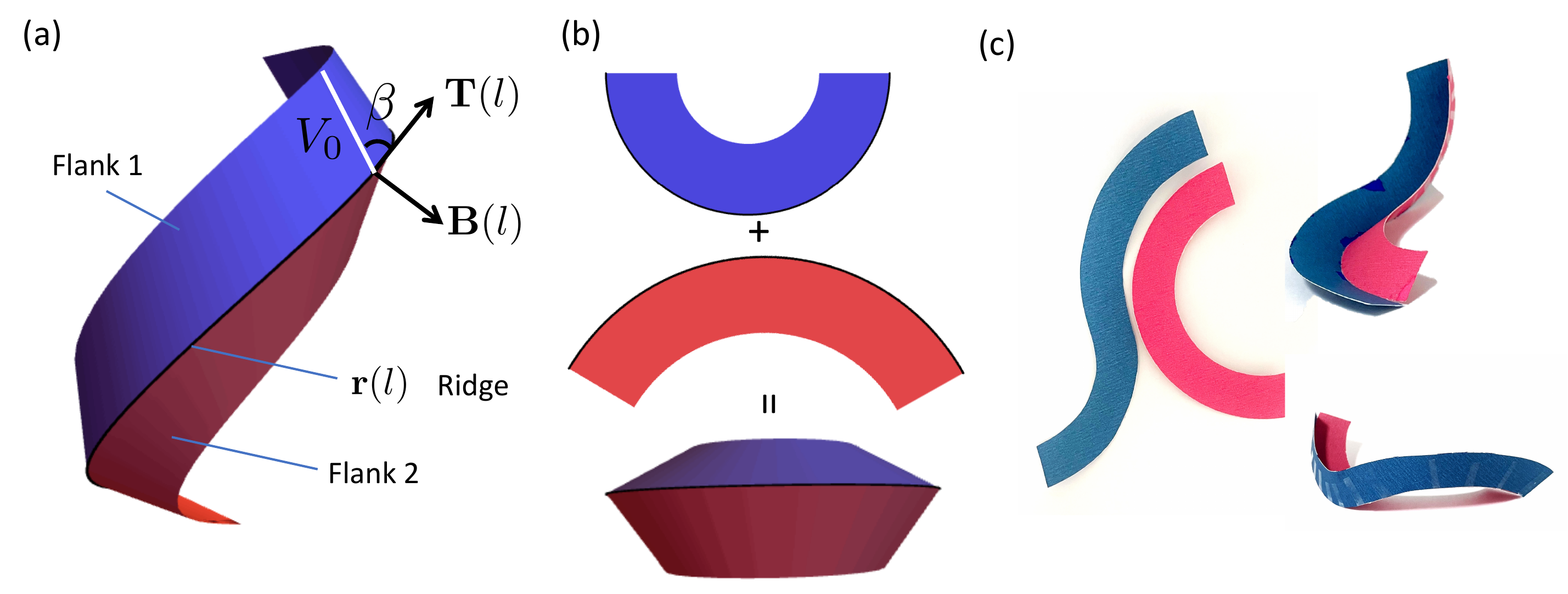}
	\caption{(a) Geometry of a curved fold. (b) Curved fold with homogeneous geodesic curvatures. (c) Generic curved fold shows non-zero torsion along the ridge.}
	\label{fig:SI_equilibrium}
\end{figure}

\section{Equilibrium configuration}
The equilibrium configuration of an ICF is determined by minimizing its total elastic energy, which, using our above results we may write as 1D integral along the ridge. During minimization, the $\kappa_{gi}$ are fixed, and the angles $\alpha_i$ and $\beta_i$ follow from the form of the ridge line via the kinematic constraints in Eqs. (\ref{eq:beta}) and (\ref{eq:alpha}). Minimization is thus only over the form of the ridge line, which, according to the fundamental theorem of space curves, is uniquely and exactly specified by the two scalar functions, torsion $\tau(l)$ and curvature $\kappa(l)$. Using our above expressions for the energies as 1D integrals along the ridge line, we thus write the minimization problem as
\beq
E=\min _{ \substack{\tau(l),\mathrm{\ }\kappa(l)\mathrm{,\ s.t.} \\ \kappa\geq\mathrm{max}(|\kappa_{g_1}|,|\kappa_{g_2}|)  }}\int_0^{L_0}   \left(\rho_r(\kappa)+ \rho_{b_1}(\kappa,\kappa', \tau)+\rho_{b_2}(\kappa,\kappa', \tau) \right)\mathrm{d} l. \label{eq:total_energy}
\eeq
In the above expression, $\rho_{b_i}$ is the flank bending energy per unit arclength of the ridge, i.e. the integrand of Eq. (\ref{eq:bending_energy}), but with $\beta$ substituted for $\kappa$, $\tau$ and $\kappa'$ using Eqs. (\ref{eq:beta}) and (\ref{eq:alpha}). The ridge energy, $\rho_r$ could take many forms depending on the situation, but generically only depends on $\kappa$, as this  determines the fold angle. For example, for a freely jointed hinge we have $\rho_r=0$, a sprung hinge with preferred angle $\gamma$ has $\rho_r(\kappa) = 1/2(\alpha_1-\alpha_2-\gamma)^2 = 1/2(\cos^{-1}(\kappa_{g_1}/\kappa)+\cos^{-1}(\kappa_{g_2}/\kappa)-\gamma)^2$, and a metric mechanics sheet has the form given in the main text, $\rho_r=\frac{1}{4\sqrt{6}} \mu t^{5/2} |\boldsymbol{\kappa}|^{-3/2}(\kappa_{g_1} + \kappa_{g_2})^2$.

The energy in Eq.\ (\ref{eq:total_energy}) can be minimized over $\tau$ and $\kappa$ variationally. Minimization over $\tau$ yields a cumbersome algebraic equation, allowing one to express $\tau$ as a local algebraic combination of $\kappa$ and $\kappa'$. Substituting this back in, minimization over $\kappa$ yields a  second order Euler-Lagrange equation, $g(\kappa(l),\kappa'(l), \kappa''(l))=0$,  that must be augmented by two suitable boundary conditions, such as the fold angle at either end of the ICF. Here, we do not attempt a full treatment of this minimization problem, but instead consider some important special cases.

%Generically the equilibrium configuration of an ICF is determined by minimizing the total elastic energy $\int_0^{L_0} \rho_E \d l = \int_0^{L_0} \rho_r(\alpha_1,\alpha_2) \d l + \frac{D}{2} \int_0^{L_0} \rho_{b_1} \d l + \frac{D}{2} \int_0^{L_0} \rho_{b_2} \d l$ over the shapes of flanks, where the ridge energy $\rho_r(\alpha_1,\alpha_2)$ is usually a function of opening angles and $\rho_{b_i}$ is the flank bending energy per unit arclength of the ridge.
%We notice that $\alpha_i$ is a function of $\kappa$ and $\rho_{b_i}$ are functions of $\kappa$ and $\tau$ under isometric deforamtions, which are determined by the shape of the ridge $\bfr(l)$. Thus we may minimize the total energy over the shape of the ridge instead, or equivalently over $\kappa$ and $\tau$, to determine the equilibrium configuration. We discuss  three cases below according to the geodesic curvatures of the reference strips.

{\bf Case 1: homogeneous geodesic curvature ($\kappa_{g_1}=const$, $\kappa_{g_2}=const$)}.  Owing to the translational invarience of the ICF, in this case it is reasonable to search for similarly invarient minimizers, with constant $\tau$ and $\kappa$, i.e.\ helices. Minimizing over $\tau$, we now find $\tau=0$, i.e. the ICF forms a plane curve.  Substituting this into $\beta$, we find $\beta=\pi/2$, i.e., the generator is perpendicular to the ridge, and thus the length of generator is identical to the width of the strip ($V_0 = w$).

Subsituting this forms for $\tau$ and $\beta$ into the energy, we find the total energy density $\rho_E$ (per unit length of the ridge) as
\begin{align}
\rho_{\rm E} =  \underbrace{\rho_r(\kappa)}_{\text{ridge energy}} + \underbrace{{\frac{1}{6}} \mu t^3 w_1 \frac{\kappa^2 - \kappa_{g_1}^2}{-\kappa_{g_1}} \log (1 - \kappa_{g_1} w_1)}_{\text{bending energy of the left flank}}
+ \underbrace{{\frac{1}{6}} \mu t^3 w_2 \frac{\kappa^2 - \kappa_{g_2}^2}{-\kappa_{g_2}} \log (1 - \kappa_{g_2} w_2)}_{\text{bending energy of the right flank}}.
\label{eq:homogeneous}
\end{align}
Minimization over $\kappa$ depends on the form of the ridge energy. We consider two special cases: 
\begin{enumerate}
    \item Freely joined folds. The ridge energy of freely joined folds vanishes, i.e., $\rho_r(\kappa)=0$. Minimizing (\ref{eq:homogeneous}) over $\kappa$ under the geometric constraint $\kappa \geq |\kappa_{g_i}|$, we obtain
    \begin{align}
    \kappa = \max \{|\kappa_{g_1}|,|\kappa_{g_2}| \},
    \end{align}
    meaning that the fold favors a small curvature and hence a small opening angle according to the geometric relation $\alpha_i = \arccos(\kappa_{g_i}/\kappa)$. 

    \item Metric-mechanic folds. In the metric-mechanic folds using LCE sheets, the effective ridge energy is given by 
\begin{align}
\rho_r=\frac{1}{4\sqrt{6}} \mu t^{5/2} |\boldsymbol{\kappa}|^{-3/2}(\kappa_{g_1} + \kappa_{g_2})^2.
\label{eq:SI_ridge_density}
\end{align}
Minimizing the total energy including the ridge energy and the flank bending energy yields the equilibrium $\kappa$ as 
\begin{align}
\kappa= \frac{3^\frac{3}{7}}{2} t^{-\frac{1}{7}}(\kappa_{g_1}+\kappa_{g_2})^\frac{4}{7} \left( \frac{\log(1-\kappa_{g_1}w_1)}{\kappa_{g_1}}+ \frac{\log(1-\kappa_{g_2} w_2)}{\kappa_{g_2}} \right)^{-\frac{2}{7}}, \label{eq:eq_curve}
\end{align}
or in the narrow ribbon limit,
\beq
\kappa =(3^3/2^5)^{1/7} \kappa_g^{4/7} t^{-1/7} w^{-2/7},
\label{eq:SI_simplified_curvature}
\eeq
as discussed in the main text.
\end{enumerate}
In both cases, these translationally invarient states correspond to global minimizers, but other in-homogeneous morphologies may be obtained by applying non-minimizing boundary conditions on the ICF --- in which case, the energy must be minimized variationally.

{\bf Case 2: symmetric inhomogeneous geodesic curvature ($\kappa_{g_1}(l) = \kappa_{g_2}(l) \equiv \kappa_{g}(l) \neq const$)}. Owing to the difficulty of solving the original Wunderlich functional, we compute the equilibrium state of the symmetric case in the narrow ribbon limit ($w_1=w_2=w \ll 1/\kappa_g$). The total  elastic energy per unit arclength of the ridge is given by
\begin{align}
\rho_{\rm E}= \rho_r(\kappa) + \frac{1}{6} \mu t^3 w \frac{\kappa^2 - \kappa_{g}^2}{\sin^4 \beta_1} 
+ \frac{1}{6} \mu t^3 w \frac{\kappa^2 - \kappa_{g}^2} {\sin^4\beta_2},
\label{eq:asymmetric}
\end{align} 
where $w$ is the width of the strip. Recalling the sign convention in Sec.~\ref{sec:SI_isometry}, we have  $\alpha_1=\arccos(\kappa_{g_i}(l)/|\boldsymbol{\kappa}(l)|)\equiv \alpha$ and  $\alpha_2=-\alpha$. The expressions for the $\sin^4\beta_i$ are then given by
\begin{align}
    &\sin^4 \beta_1=\kappa^4 \sin^4\alpha/\left[(\tau-\alpha')^2+\kappa^2\sin^2\alpha\right]^2, \\ \nonumber
  &\sin^4\beta_2=\kappa^4 \sin^4\alpha/\left[(\tau+\alpha')^2+\kappa^2\sin^2\alpha \right]^2,
\end{align}
In this case, minimization over $\tau$ gives $\tau=0$ due to the symmetry of the two $\tau$ containing terms. Minimization over $\kappa$ then depends on the form of the ridge energy. Again we consider two special cases:
\begin{enumerate}
    \item Freely joined folds, $\rho_r(\kappa)=0$. The global minimizer of $\rho_E$ is  $\kappa(l) = \kappa_{g_i}(l)$, as then $\rho_E=0$, which must be the minimum given $\rho_E$ is strictly positive. This corresponds to the zero energy closed-book state on a plane, with zero ridge energy or bend energy. Other states may be observed by imposing non-minimizing boundary conditions, but the ICF will always be torsion-free.

    \item Metric-mechanic folds.  In the metric-mechanic folds using LCE sheets, the effective ridge energy is again given by 
\begin{align}
\rho_r=\frac{1}{4\sqrt{6}} \mu t^{5/2} |\boldsymbol{\kappa}|^{-3/2}(\kappa_{g_1} + \kappa_{g_2})^2.
    \end{align}
    Minimization over $\kappa$ yields an ODE $g(\kappa(l),\kappa'(l), \kappa''(l))=0$ that will determine the shape of the ridge by solving corresponding boundary value problems. 
\end{enumerate}

%{\bf Case 3: generic case}. As shown in Fig.~5 of the main text and also in Fig.~\ref{fig:SI_equilibrium}(c), the equilibrium state of a fold composed of two generic strips apart from the first two cases will possess non-zero torsion along the ridge. However, the problem of determining the equilibrium shape of a generic fold remains open and we reserve it as a future direction. Moreover, the more interesting inverse problem, which involves identifying two reference strips capable of achieving a target curved fold in 3D, would benefit engineering applications such as actuators and robotic arms.  

\section{Designing ICFs via metric mechanics}
\begin{figure}[!h]
	\centering
	\includegraphics[width=0.8\columnwidth]{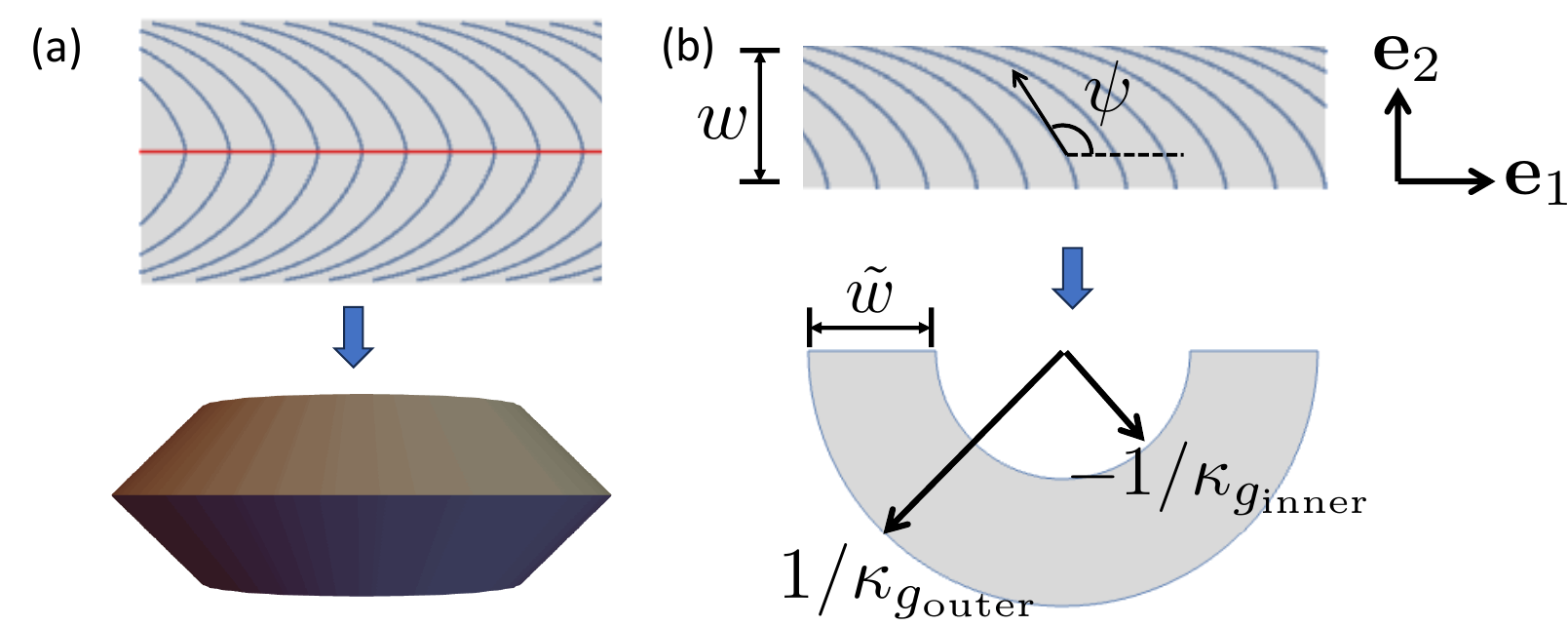}
	\caption{(a) A translationally invariant pattern morphs into a surface of revolution with a sharp ridge following the exact metric. (b) An individual pattern morphs into an arc strip.}
	\label{fig:SI_pattern}
\end{figure}

A 2D LCE sheet with a patterned director field $\bfn(x,y)$ will morph into a curved surface in 3D via heating from nematic state to isotropic state. Locally, the sheet contracts by a factor $\lambda_{\parallel}$ along the director and elongates by a factor $\lambda_{\perp}$ perpendicular to it, inducing a metric tensor 
\begin{equation}
\bar{a}(x,y) = \lambda_{\parallel}^2 \bfn(x,y) \otimes\bfn(x,y) + \lambda_{\perp}^2 \otimes \bfn^\perp(x,y) \bfn^\perp(x,y).
\end{equation}
As discussed in the main text, here we seek to design a director pattern with translational symmetry that will morph into a homogeneous ICF, i.e., a surface of revolution with a shape ridge (Fig.~\ref{fig:SI_pattern}(a)).
Following the main text, we first seek a pattern that will morph a strip into an arc. We thus consider an initially flat strip occupying $0<y<w$ in with director pattern $\bfn(x,y) = \cos\psi(x,y) \bfe_1 + \sin \psi(x,y) \bfe_2$, as shown in Fig.\ \ref{fig:SI_pattern}(b). 
A direct application of the \emph{Theorema Egregium} \cite{o2006elementary} shows that the Gaussian curvature of the strip after actuation will be 
\cite{aharoni2014geometry, mostajeran2015curvature}:
\beq
K=(\lambda_{\parallel}^{2} -\lambda_{\perp}^{2})\left[\cos(2\psi)\left(\left(\frac{\partial\psi}{\partial y}\right)^2 - \left(\frac{\partial\psi}{\partial x}\right)^2 +\frac{\partial^2\psi}{\partial x \partial y}\right)+ \frac{1}{2}\sin(2\psi)\left(\frac{\partial^2\psi}{\partial x^2}-4 \frac{\partial   \psi}{\partial x} \frac{\partial \psi}{\partial y}+ \frac{\partial^2\psi}{\partial y^2}\right)\right]. \label{eq:gauss}
\eeq
A similar direct computation \cite{duffy2020defective, feng2022interfacial} shows that a translationally-invariant ($\psi(x,y)=\psi(y)$) straight line in the $\bfe_1$ (length) direction will, after actuation, have geodesic curvature
\begin{align}
\kappa_g =\frac{-\left(\frac{\lambda_{\perp}}{\lambda_{\parallel}} + \frac{\lambda_{\parallel}}{\lambda_{\perp}} \right)\frac{\d \psi}{\d y} \sin\psi \cos \psi}{\sqrt{\lambda_{\parallel}^2 \cos^2\psi + \lambda_{\perp}^2\sin^2\psi}},\label{eq:si_geodesic}
\end{align}
where $\psi(y)$ is the director angle along the line in the reference state.

We search for a translationally invariant pattern $\psi(x,y)=\psi(y)$ such that the actuated state is an arc-like strip in 2D. We thus set $K=0$, so that the Gauss equation reduces to $\cos(2\psi) (\partial y  \psi)^2 + (1/2) \sin(2\psi)\partial y \partial y  \psi =0$. Solving, we obtain
\beq
\psi(y)=\frac{1}{2} \arccos(c_1  y + c_2)
\eeq
where $c_1$ and $c_2$ are constants of integration. To achieve the maximal actuation, we  choose the constants $c_1$, $c_2$ so that the director is parallel and perpendicular to the lower and upper boundaries, leading to the form:
\beq
\psi(y) = \frac{1}{2}\arccos(2y/w-1).
\eeq
Substituting this into Eq.\ \ref{eq:si_geodesic}, the geodesic curvatures of the boundaries following the sign convention in Fig.~\ref{fig:SI_isometry} and Fig. 2 are
\beq
\kappa_{g_\mathrm{inner}} =\frac{\lambda_{\parallel}^2 -  \lambda_{\perp}^2}{2 w \lambda_{\parallel}^2 \lambda_{\perp}}, \,\,\,  \kappa_{g_\mathrm{outer}}= \frac{-\lambda_{\parallel}^2 +  \lambda_{\perp}^2}{2 w \lambda_{\parallel} \lambda_{\perp}^2}.
\eeq
Since these are uniform, the pattern thus does morph the strip into an arc, with inner and outer radii $-1/\kappa_{g_\mathrm{inner}}$ and $1/\kappa_{g_\mathrm{outer}}$, as shown in Fig.\ \ref{fig:SI_pattern}. Accordingly, the width of the strip after actuation is $\bar{w} = 1/\kappa_{g_{\text{inner}}} + 1/\kappa_{g_{\text{outer}}}$. 

\begin{figure}[!h]
	\centering
	\includegraphics[width=\columnwidth]{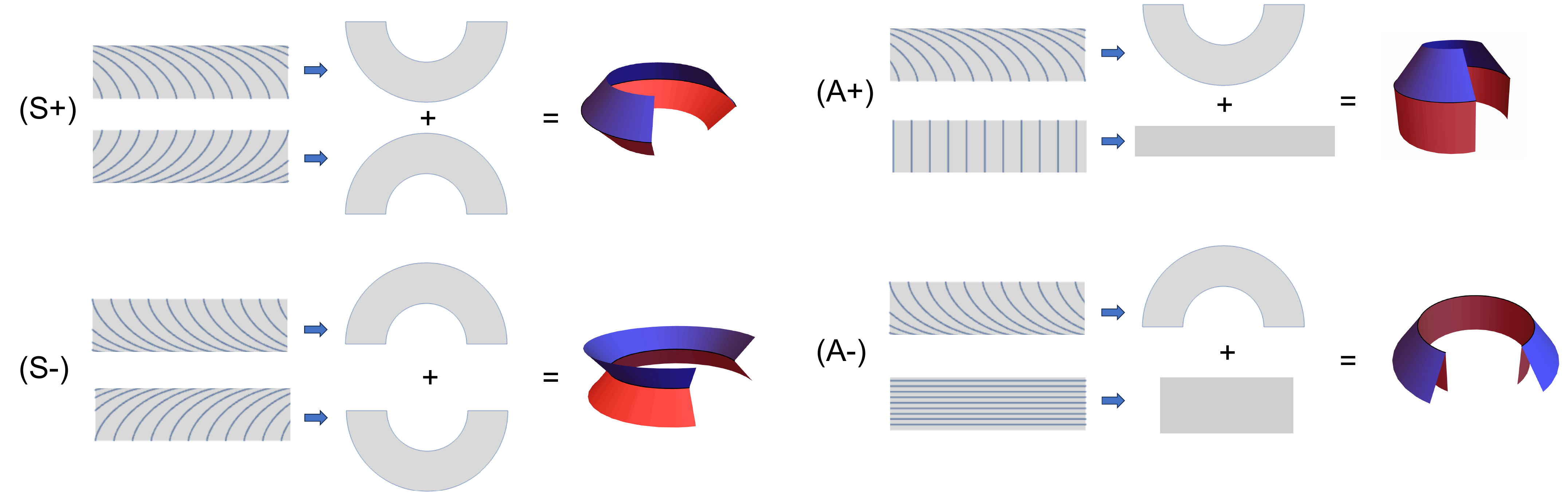}
	\caption{Constructions of four types of ICFs.}
	\label{fig:SI_fourfolds}
\end{figure}

As shown in the main text and Fig.~\ref{fig:SI_fourfolds}, stitching together two strips in different ways will result in four types of ICFs, classified as symmetric positive (S+), symmetric negative (S-), asymmetric positive (A+) and asymmetric negative (A-), based on the concentrated GC and the symmetry. Following the sign convention in the main text, the concentrated GCs, i.e. $\kappa_{g_1} + \kappa_{g_2}$, are $2\kappa_{g_{\text{outer}}}$ (S+), $2\kappa_{g_{\text{inner}}}$ (S-), $\kappa_{g_{\text{outer}}}$ (A+) and $\kappa_{g_{\text{inner}}}$ (A-), respectively. In each case, the stitching trivially obeys the basic rule of metric compatibility, as the director is continuous across the boundary, so the two pattern regions agree on the interface's length. The patterns can also be joined in additional ways, by  taking interfaces with a discontinuous director, provided the interfaces bisect  the directors on either side \cite{feng2022interfacial}.

% As shown in the main text and Fig.~\ref{fig:SI_fourfolds}, stitching together two strips in different ways will result in four types of ICFs, classified as symmetric positive (S+), symmetric negative (S-), asymmetric positive (A+) and asymmetric negative (A-), based on the concentrated GC and the symmetry. Following the sign convention in the main text, the concentrated GCs, i.e. $\kappa_{g_1} + \kappa_{g_2}$, are $2\kappa_{g_{\text{outer}}}$ (S+), $-2\kappa_{g_{\text{inner}}}$ (S-), $\kappa_{g_{\text{outer}}}$ (A+) and $-\kappa_{g_{\text{inner}}}$ (A-), respectively.

\section{Ridge energy for LCE sheets}
Following the exact metric,  the metric-mechanic ICF will form a sharp ridge after actuation. However, this configuration is not physically observed, as such a sharp ridge would have divergent transverse curvature and hence infinite bending energy. Instead, a relaxed shape with a smooth ridge is observed in experiments and simulations, which emerges via energy minimization involving a stretch-bend trade-off.  Here we quantify this ridge energy for LCE sheets. Our result can then be minimized, along with the flank bending energies, to predict the observed form of an unloaded LCE ICF. The code related to this section is released at GitHub \url{https://github.com/fengfan628/intrisically_curved_folds} with a CC-BY-4.0 license.

\subsection{Exact energy for ridge relaxation} 

\begin{figure}[!h]
	\centering
	\includegraphics[width=\columnwidth]{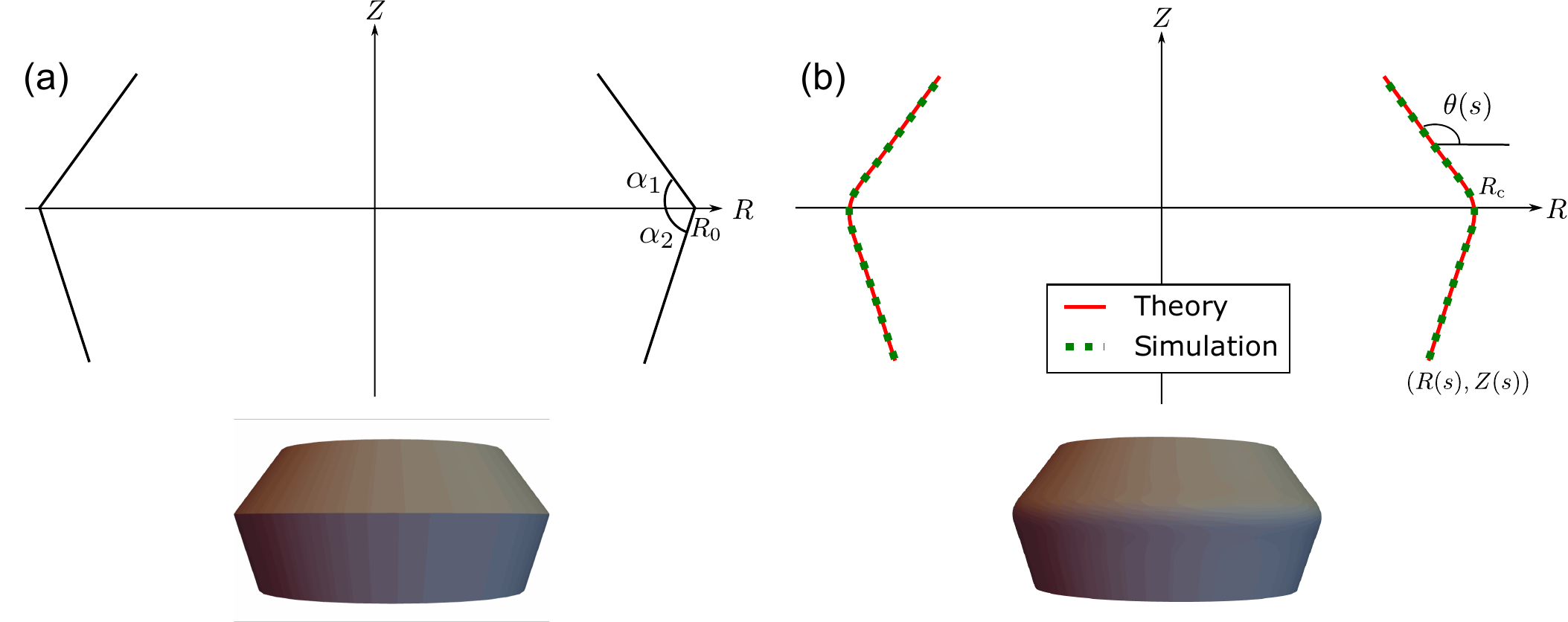}
	\caption{(a) $(R,Z)$ slice of the initial surface of revolution with a sharp ridge. (b) Blunted ridge $(R(s), Z(s))$ after energy minimization. The parameters are: $R_0=1$, $t=0.01$, $\alpha_1=0.3\pi$, $\alpha_2=0.4\pi$.}
	\label{fig:SI_relaxation}
\end{figure}

We start by considering a homogeneous ICF that follows an exact isometry with curvature $\kappa$. Such an ICF will form a surface of revolution consisting of two conical flanks meeting at a sharp ridge with radius $R_0=1/\kappa$. We show such an $(R,Z)$ cross-section of such a shape in  Fig.~\ref{fig:SI_relaxation}, alongside the actual surface of revolution, that is generated by rotating the 2D wedge-like cross-section about the axis $Z$. As ever, the fold angles $\alpha_i$ follow kinematically from the $\kappa$ and the $\kappa_{gi}$ that define the fold. In the $(R,Z)$ plane, this reference curve is given by
\begin{align}
  \bar{R}(s) = \begin{cases}
      R_0 -s \cos\alpha_1, \quad s\geq0\\
      R_0 + s \cos\alpha_2, \quad s<0
  \end{cases}, \quad
    \bar{Z}(s) = 
      s \sin\alpha_1
\end{align}
where $s$ is the arclength of the curve.
The reference configuration possesses a sharp ridge at $s=0$ when $\alpha_1 + \alpha_2 \neq \pi$.

To compute the analytical form $(R(s), Z(s))$ of the blunted ridge by energy minimization, we consider the surface as
 a thin sheet of incompressible Neo-Hookean elastomer with thickness $t$. The total elastic energy $E$ contains two contributions, the bending energy $E_{b}$ and the stretching energy $E_{s}$, with the forms \cite{duffy2021shape}
\beq
E_{s} = \iint \frac{\mu t}{2} \left(\lambda_1^2 + \lambda_2^2 + \frac{1}{\lambda_1^2 \lambda_2^2} - 3 \right) \d S \label{eq:stretch_energy}
\eeq
and
\beq
E_{b}= \iint \frac{\mu t^3}{6} (\kappa_1^2 +\kappa_2^2+\kappa_1 \kappa_2)  \d S 
\label{eq:bend_energy}
\eeq
where $\mu$ is the shear modulus, $\lambda_i$ are the (in-plane) principal stretches away from the exact isometry, and  $\kappa_i$ are the principal curvatures of the surface. Normally, the higher thickness scaling of the bending energy makes it negligible compared to stretch, so that the stretch energy effectively becomes a constraint limiting consideration to isometric deformations. In such cases, the bend acts only as a tie break, choosing the bend-energy minimizing isometric deformations. However, here the available isometry has divergent bend, so a non-trivial trade-off between the two must occur. 

 In the case of a homogeneous ICF, both the reference and relaxed shapes are surfaces of revolution, whose shapes can be described by the slice $(R(s),Z(s))$ in 2D. Accordingly, the principal stretches may be given by
 \beq
 \lambda_1= R(s)/\bar{R}(s),\quad \lambda_2=\sqrt{R'(s)^2 + Z'(s)^2}
 \label{eq:stretch}
 \eeq
and the principal curvatures given by
\beq
\kappa_1=\frac{-R'' Z' + R' Z''}{\sqrt{R'^2 + Z'^2}},\quad \kappa_2=\frac{Z'/R}{\sqrt{R'^2 + Z'^2}}. 
\label{eq:bend}
\eeq
Substituting these forms into $E_s$ and $E_b$, one obtains an energy that is a simple functional of $R(s)$, $Z(s)$ and their first and second derivatives.

\subsection{Full numerical relaxation} 
The most direct route to finding the shape of the ridge is then to minimize this energy numerically, 
which we do in Python (also in Mathematica) using a simple 1D finite element scheme. The 1D reference and relaxed curves in the $(R,Z)$ plane are discretized correspondingly. The resulting stretches (\ref{eq:stretch}) and principal curvatures (\ref{eq:bend}) are  computed numerically in a finite element scheme. Then we use the BFGS method in scipy to minimize the total elastic energy $E_s + E_b$ in (\ref{eq:stretch_energy}) and (\ref{eq:bend_energy}), which leads to the relaxed shape.
See GitHub \url{https://github.com/fengfan628/intrisically_curved_folds} for more detail. An example of such a solution is the ``simulation" curve in Figs. \ref{fig:SI_relaxation}(b), which, as expected, shows the sharp crease relaxing into a smooth structure.

\subsection{Analytic relaxation in thin and nearly cylindrical limit}

Our simulations show that blunting is confined to a small region around the ridge, that, as expected, becomes ever smaller in the thin limit, as the ridge comes closer to the sharp true isometry.  To obtain further theoretical insight, we use this fact, and our simulations, to make several approximations that then allow a simple analytic solution for the ridge shape.  Post-hoc, we see that the approximations are self-consistent in the thin limit, $t \ll R_0$. 

Firstly, we assert that  the strains are geometrically small (though the rotations may be large), that is, $\lambda_1 = 1 + \epsilon_1$ and $\lambda_2 = 1 + \epsilon_2$ with $\epsilon_1 \ll 1$ and $\epsilon_2 \ll 1$. The stretching energy may then be expanded in terms of $\epsilon_1$ and $\epsilon_2$ as
\beq
E_{s} = \frac{\mu t}{2} (4 \epsilon_1^2 + 4 \epsilon_2^2 + 4 \epsilon_1 \epsilon_2) + \text{higher order terms}.
\eeq
Secondly, we assume the elastomer is able to relax to its minimizing strain in the generator direction, leading to $\epsilon_2 = -\epsilon_1/2$. This assumption is equivalent to assuming the azimuthal hoop stress dominates the $ss$ stress, which is well justified by the simulation results, and self-consistently correct in the subsequent analysis. This second approximation reduces the stretching energy to
\beq
{E_{\rm s} =\frac{\mu t}{2} (3 \epsilon_1^2)} =\frac{E t}{2} \epsilon_1^2 =\frac{Y}{2} \epsilon_1^2 =\frac{Y}{2}\left( \frac{\Delta R(s)}{\bar{R}(s)}\right)^2 .
\eeq
where $E=3 \mu$ is the Young's modulus of the LCE, which is the natural modulus for uniaxial stretches, and $E t=Y$ is the stretching modulus of the sheet as used in the main text. 

Thirdly, we assume the bending energy is dominated by the generator bend, $\kappa_1$ as this is what diverges in the sharp limit. Defining the angle $\theta(s)$ as the angle between the relaxed curve and the radial direction (see Fig.\ \ref{fig:SI_relaxation}B), then, as strains are small, we simply have $\kappa_1=\theta'(s)$, so the dominant bending energy is:
\begin{equation}
E_{\rm b} =\frac{1}{2} D \theta'(s)^2.
\end{equation}
Combining these, leads to the total energy
\begin{equation}
E=\int \left(\frac{Y}{2}\left( \frac{\Delta R(s)}{\bar{R}(s)}\right)^2 +\frac{1}{2} D \theta'(s)^2\right) 2 \pi \bar{R}(s) \mathrm{d}s.
\end{equation}
To make further progress, we fourthly assume $R_0$ is large compared to the extent of the blunted crease, so that we may neglect the distinction between $\bar{R}(s)$ and $R_0$ in the region of interest:
\begin{equation}
E=2 \pi R_0  \int \left(\frac{Y}{2}\left( \frac{\Delta R(s)}{R_0}\right)^2 +\frac{1}{2} D \theta'(s)^2\right) \mathrm{d}s.\label{eq:energy_cylindrical}
\end{equation}
Finally, we assume that $\alpha_i \approx \pi/2$, i.e. the surface of revolution is nearly a cylinder, which simplifies the curvature to $\kappa_1=\theta'(s) \approx \Delta R ''(s)$. This approximation is the least well justified, as the $\alpha_i$ are determined from the curvature by kinematics, and, in general, only become nearly cylindrical in the highly curved configuration. Nevertheless, making this approximation, we have the simple energy
\begin{equation}
E=2 \pi R_0  \int \left(\frac{Y}{2}\left( \frac{\Delta R(s)}{R_0}\right)^2 +\frac{1}{2} D \Delta R''(s)^2\right) \mathrm{d}s.
\end{equation}
Minimizing variationally over $\Delta R(s)$ yields the Euler-Lagrange equation as
\beq
\Delta R^{(4)}(s) + \frac{Y}{D R_0^2} \Delta R(s) = 0, \label{eq:el_cylindrical}
\eeq
which admits four independent solutions $ \Delta R(s)\propto \exp((\pm 1 \pm i)s/f)$, revealing 
$f=(4 R_0^2D/Y)^{1/4} \propto \sqrt{R_0 t}$ as the emergent length-scale of the blunted domain, where $Y=3\mu t$ and $D=\frac{1}{3}\mu t^3$. Taking the decaying solutions at infinity, we have
\begin{align}
   \Delta R(s) = \begin{cases}
       c_1 \exp(-s(1-i)/f) + c_2 \exp(-s(1+i)/f) \quad  s>0 \\
       c_3 \exp(s(1-i)/f) + c_4 \exp(s(1+i)/f) \quad  s\leq0
   \end{cases}
\end{align}
Joining them at $s=0$ under the conditions $\Delta R(0) = d_1$ and $\Delta R'(0) = d_2$ and then minimizing the total energy $E$ over $(d_1,d_2)$, we have
\begin{align}
\Delta R(s)=  - \frac{f}{4} \exp \left(-|s|/f\right) (\cos\alpha_1 + \cos\alpha_2) \left(\cos(s/f) - \sin(|s|/f)\right)  \label{eq:shape}.
\end{align}
The actual form of the surface is then reconstructed as  $R(s) = \bar{R}(s) + \Delta R(s)$, and $Z(s)=\int \sqrt{1-R'(s)^2}\mathrm{d}s$. Despite the numerous approximations, we see in Fig.\ \ref{fig:SI_relaxation} that this solution agrees almost perfectly with the full numerical solution. 

\begin{figure}[!h]
	\centering
	\includegraphics[width=\columnwidth]{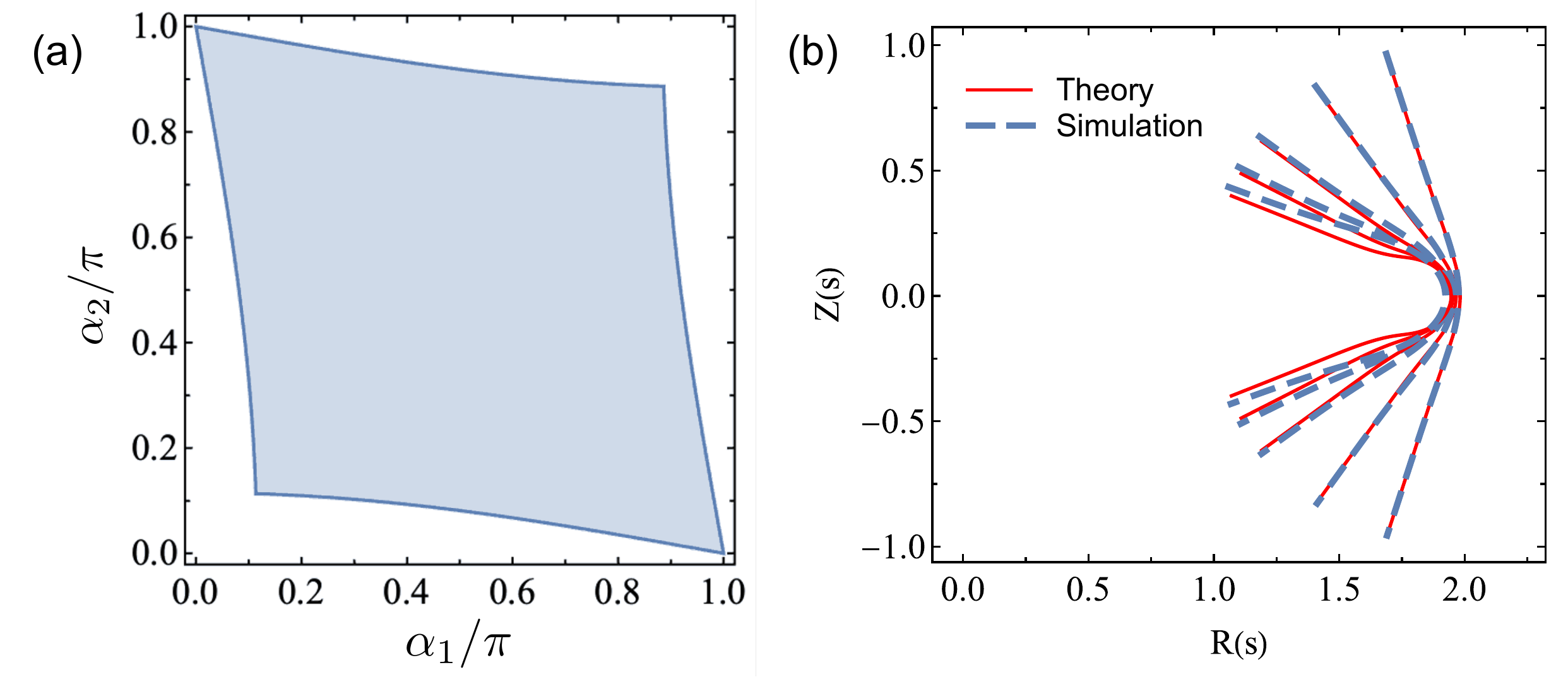}
	\caption{(a) Admissible domain of $(\alpha_1,\alpha_2)$. (b) Theoretical (Eq. \ref{eq:shape}) and simulated relaxed shapes for $\alpha_1=\alpha_2 = (0.12,0.15,0.2,0.3,0.4)\pi$.}
	\label{fig:s_adm}
\end{figure}

Substituting the relaxed shape (\ref{eq:shape}), we obtain the effective ridge energy density $\rho_r = E_r/(2\pi R_0)$ as
\begin{align}
\rho_{r}&=\frac{1}{4\sqrt{6}} \mu t^{5/2} R_0^{-1/2}(\cos\alpha_1 + \cos\alpha_2)^2 \nonumber\\
&=\frac{1}{4\sqrt{6}} \mu t^{5/2} |\boldsymbol{\kappa}|^{-3/2}(\kappa_{g_1} + \kappa_{g_2})^2.
\label{eq:s_ridge_density}
\end{align}
Having completed this calculation, we observe that both the length of the blunting region and the associated size of $\Delta R$ both scale as $\sqrt{R_0 t}$, which is an intermediate lengthscale, short compared to $R_0$ but long compared to $t$. Post-hoc, this intermediate scale justifies  all our  assumptions, ensuring that our analytic solution is asymptotically exact in the thin limit, $R_0 \gg t$. Precisely, the length of the blunted crease is $\sim \sqrt{R_0 t} \ll R_0$, justifying the approximation $\bar{R}(s)=R_0$ throughout the region of interest. Furthermore, the  characteristic size of hoop strain is  $\Delta R/R_0 \sim \sqrt{t/R_0} \ll 1 $ which vanishes in the thin limit, justifying the small strain approximation. Correspondingly, the azimuthal membrane stress has scale $N{\phi \phi}=E t (\Delta R/R_0) \sim E t^{3/2}/R^{1/2}$. In contrast, the leading generator stress can be found from vertical force balance as $N{ss} \sim D \theta''(s) \sim D /f^2\sim E t^2/R_0$. Hence the $ss$ stress is smaller than the hoop stress by a scaling factor of $\sqrt{t/R_0}$, confirming it is negligible in the thin limit, justifying the approximation $\epsilon_2=-\nu \epsilon_1$.

Such considerations show that all the key assumptions are justified in the thin limit, except the nearly cylindrical one, $\alpha_i \approx \pi/2$. The equilibrium configuration of a free-floating ICF is given by minimizing total energy consisting of this ridge energy (Eq.\ \ref{eq:s_ridge_density}) and the bending energy of the flanks (Eq.\ \ref{eq:bending_energy}). This minimization gives the $\kappa \propto t^{-1/7}$ (Eq.\ \ref{eq:eq_curve}), showing that the curvature itself diverges in the thin limit, so, via the kinematic relations, the fold will be almost cylindrical. Thus all the approximations are valid in the thin limit for the equilibrium configuration, explaining the good agreement between theory and numerics/experiment. However, in other circumstances (e.g. under load) one may have a thin ICF that is not nearly cylindrical, so it is useful to probe the range of accuracy/validity of this final approximation. At the extreme limit, the solution is only well defined if $|R'(s)|<1$, reducing the admissible domain of $\alpha_1$ and $\alpha_2$ to that shown in fig.~\ref{fig:s_adm}. Within this region, we compare the approximate and full numerical solutions for positive symmetric folds with $\alpha_1=\alpha_2\in(0,\pi/2)$ in Fig.~\ref{fig:s_adm}(b), revealing that a good fit is achieved over a surprisingly wide region of folds, even far from the cylindrical limit, with  deviations only becoming significant when  $\alpha_1=\alpha_2 < 0.2 \pi$.

\subsection{Analytic shape equation for far-from-cylindrical ICFs}

\begin{figure}[!h]
	\centering
	\includegraphics[width=0.8\columnwidth]{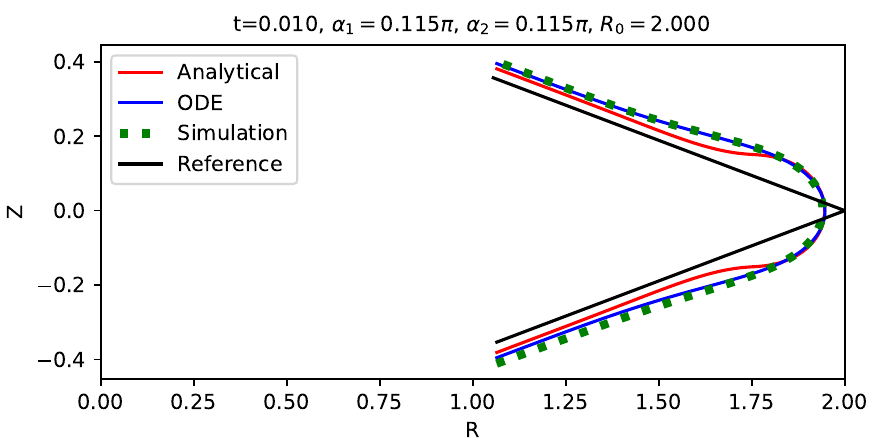}
	\caption{Relaxed shapes for the ICF with $t=0.01,\alpha_1=\alpha_2=0.115\pi, R_0=2.0$.}
	\label{fig:SI_ODE}
\end{figure}

The nearly cylindrical approximation is only used to replace $\theta'(s) \to \Delta R''(s)$ in the ICF energy. If this final approximation is invalid, one may instead use the following small-strain relations to substitute $\Delta R$ for $\theta$:
\beq
R(s)=R_c + \int_0^s \cos\theta \d s, \quad  Z(s)=\int_0^s \sin\theta \d s.
\eeq
Making this substitution in the energy Eq.\ \ref{eq:energy_cylindrical}, leads to 
\beq
E \approx 2\pi R_0 \int_{-\infty}^{\infty} \left(\frac{Y(R_0 - R_c + \int_0^s (\cos \alpha - \cos\theta(s)) \d s)^2}{2 R_0^2} + \frac{D (\theta'(s))^2}{2} \right) \d s.
\eeq
Taking the variation of $E$ with respect to $\theta(s)$ yields the corresponding ODE as
\beq
\theta^{(4)} = \frac{Y}{DR_0^2} (-\cos\alpha + \cos\theta) \sin\theta + (\theta''^2 + 2 \theta' \theta''') \cot\theta - \theta'^2 \theta'' (1+\cos^2\theta)  \csc^2\theta \label{eq:SI_ODE}
\eeq
where $\alpha$ is the angle between the reference slice and the $R$ axis. For example in Fig.~\ref{fig:SI_relaxation}(a), $\alpha = \alpha_2$ for $s<0$ and $\alpha=\pi-\alpha_1$ for $s>0$. This shape equation can be used to describe the relaxation of thin ICFs, even far from the cylindrical limit. To solve the ODE numerically, we write the generalized vector $\bfu(s) = (\theta(s), \theta'(s), \theta''(s), \theta'''(s), R(s), Z(s))$ and establish the ODE system $\bfu'(s) = \bfA(s) \bfu(s)$ by substituting (\ref{eq:SI_ODE}), $R'(s)=\cos\theta(s)$, and $Z'(s)=\sin\theta(s)$. Solving the corresponding boundary value problem numerically with solvebvp function in scip of Python gives the relaxed shape of the ridge (see GitHub \url{https://github.com/fengfan628/intrisically_curved_folds}). We compare the relaxed shapes obtained by the ODE (\ref{eq:SI_ODE}), the analytical solution (\ref{eq:shape}) and the simulation in Fig.~\ref{fig:SI_ODE}. As can be seen, the relaxed shape obtained by the ODE is more accurate when the reference configuration is far from the nearly cylindrical assumption (i.e., $\alpha_i$ is small). 

\section{Experimental: Curved Folds on Paper Models}
The curved folds paper models (Figures 4,5 in main text) were constructed using 0.25 mm thick craft paper and scotch tape. The flat individual curves were cut using a Silhouette Studio digital cutter. Asymmetric negative, asymmetric positive, symmetric negative, and symmetric positive curved folds were constructed so that the shared folds had a radius of 4.05 cm and a width 1.3 cm. For the symmetric negative curved fold, two curves (inner radius = 4.05 cm, outer radius = 5.35 cm) were laid flat and the inner curves were taped together with around 18 strips of tape, which acted as hinges, resulting in a flat curve that opened into a 3D curve with the hinge was opened. Similarly, the same protocol was followed for the symmetric positive fold, where curves (outer radius = 4.05 cm, inner radius = 2.75 cm) were taped along the outer edge. Asymmetric negative combined a straight strip of paper (12.72 cm long, 1.3 cm wide) inner edge of a curved (inner radius = 4.05 cm, outer radius = 5.35 cm) paper and the asymmetric positive combined a straight strip of paper (12.72 cm long, 1.3 cm wide) with the outer edge of a curve (outer radius = 4.05 cm, inner radius = 2.75 cm). The curved fold that produced non-zero torsion (Figure 5 in main text) was created by combining a curve with a non-uniform radius of curvature to a curve with a single radius of curvature.
\newline
\indent
Lifting experiments of the symmetric curved fold paper was model (Figure 4A in main text) as accomplished by attaching tethers to the outer flanges of both sides of the model. When the tethers are pulled the flange opens and the paper model curls into itself and is capable of gripping, lifting, and moving items (a roll of tape is shown). 
\newline
\indent
Strength experiments of the four different curved folds (symmetric negative, symmetric positive, asymmetric negative, asymmetric positive, Figure 4B in main text) were achieved by attaching tethers to both ends of the hinged folds. The paper models were then hung by their top tether and increasingly large weights were attached to the bottom tether and images were taken to determine the weight thresholds where the paper models buckled out of plane.

\section{Experimental: The LCE Intrinsically Curved Folds} 
\subsection{3D printing of LCE}
\begin{figure}[!h]
	\centering
	\includegraphics[width=1.0\columnwidth]{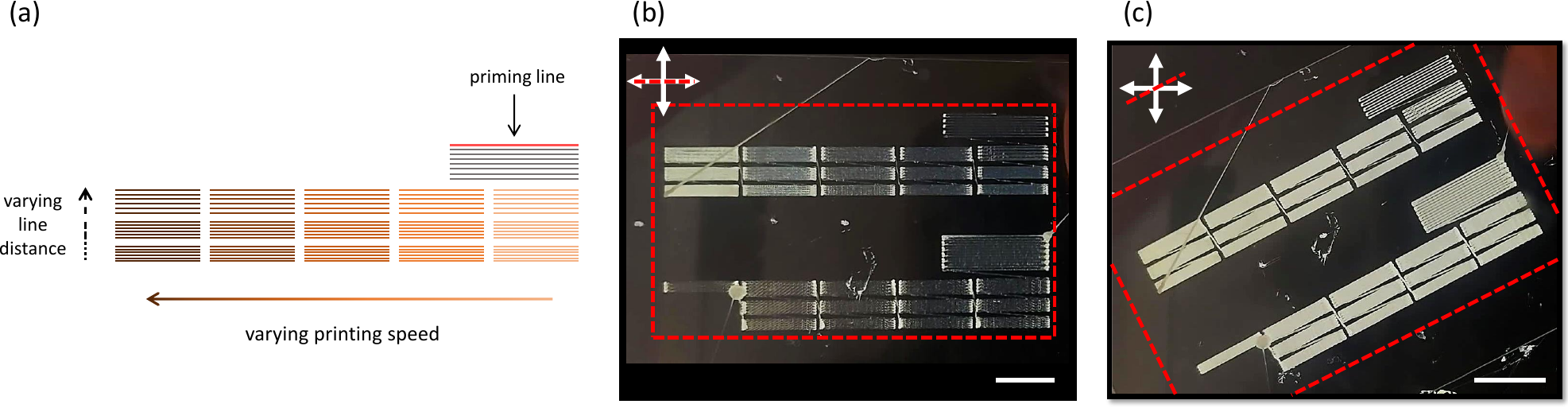}
	\caption{(a) A scheme of the calibration pattern. (b) and (c) printed calibration patterns seen through crossed polarizers while sample was rotated (white arrows in top left corner; red dotted lines present position of a glass slide against crossed polarizers). Scale bars are 1 cm.}
	\label{SI_calibration}
\end{figure}

As discussed in the main text (section: Materials and Methods), in order to print all LCE ICFs and the grippers, careful calibration of printing parameters is needed. A more detailed protocol is as follows.  First, the printing bed was leveled, and the nozzle of the syringe was placed above the PVA-coated glass such that it was touching the glass, but not pressing it. Second, the syringe with LCE ink (after oligomerization) was installed in the printer, and heated from room temperature to $80 ^\circ$C, and this temperature was maintained for 30 mins. Thirdly, to select optimal printing parameters, a calibration pattern was printed, as shown in Fig. \ref{SI_calibration}. The calibration pattern was composed of a grid of linear samples, differentiated by print speed, line separation, and priming duration (Fig. \ref{SI_calibration}(a)). Following calibration, the  printing parameters mentioned in the main text were chosen to obtain repeatable prints with good alignment. We found that print speed, extrusion rate and line separation could be held constant between batches of ink, but priming duration was reoptomised for each batch.

Print patterns were generated using Vector Slicer software for automatic slicing of director patterns. The associated code release is available under \url{https://github.com/zmmyslony/vector_slicer/releases/tag/publication}.

%number of columns with set of printing lines separated with a distance (all adjustable). Every column was printed with a speed from a speed range, with fixed distance between printed lines. This pattern was printed multiple time on glass slides with varying priming conditions. 

\label{sec:thermal_strain}

\subsection{Characterization of LCE monodomain samples}
\begin{figure}[!h]
	\centering
	\includegraphics[width=0.8 \columnwidth]{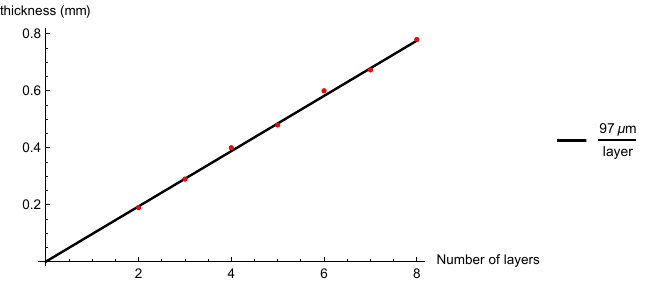}
	\caption{Thickness (unactuated) of linear patterns with different numbers of printed layers.}
	\label{Fig:SI_thickness}
\end{figure}

The thickness of linear samples with different numbers of printed layers was measured with digital calipers before actuation, showing that each layer contributed around 100 $\mu$m of thickness (Fig. \ref{Fig:SI_thickness}.)

\begin{figure}[!h]
	\centering
	\includegraphics[width= \columnwidth]{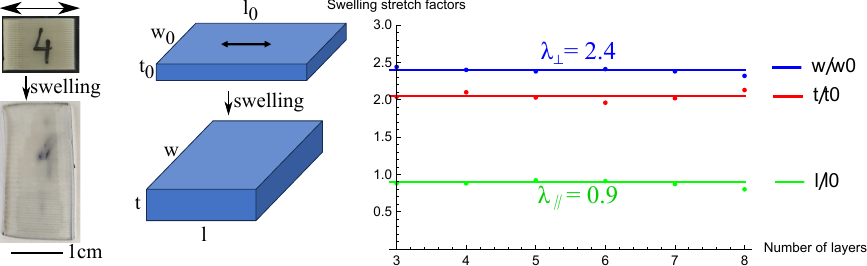}
	\caption{Swelling strains for for 3D-printed samples with linear alignment.  Left: photograph of a four layer sample before and after swelling. Middle: schematic defining the stretch ratios. Right: Swelling ratios for LCE monodomains with varying numbers of printed layers. }
	\label{swelling_ratio}
\end{figure}

The swelling ratio for printed LCEs submerged in toluene was measured for linear samples with varying thicknesses (3-8 - layers). As seen in Fig. \ref{swelling_ratio}, the in-plane swelling ratios are $\lambda_{\parallel}\approx 0.9$ and $\lambda_{\perp}\approx 2.4$, independent of thickness. Interestingly, swelling in the thickness direction itself is slightly less than $\lambda_{\perp}$, which we attribute to the corrugated nature of the print. 

\begin{figure}[!h]
	\centering
	\includegraphics[width=\columnwidth]{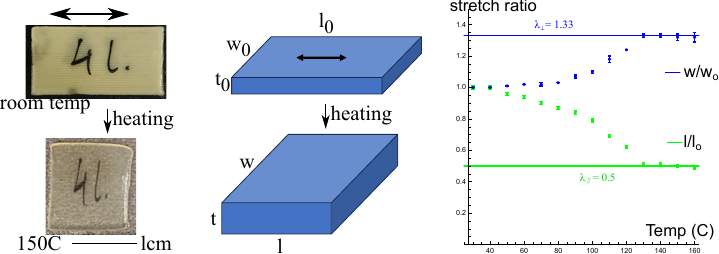}
	\caption{Thermal strain for 3D-printed samples with linear alignment. Left: photograph of 4-layer sample before and actuation on a hotplate. Middle: schematic defining the stretch ratios. Right: Planar stretch ratios as a function of temperature. Standard errors, with $n=5$ samples.}
	\label{fig:SI_linear}
\end{figure}

The thermal strain of an LCE 3D-printed sample with linear alignment was characterized on a hot plate. Five samples with planar dimensions 2 cm $\times$ 1 cm and four printed layers were placed simultaneously on the hot plate, and elongation parallel and perpendicular to alignment (printing direction) was measured  5 min after the temperature was set. The resulting data are presented in Fig. \ref{fig:SI_linear} (with standard error bars), showing that  spontaneous stretches were $\lambda_{\parallel}=0.5$ and $\lambda_{\perp}=1.33$ by the isotropic state. 


% \subsection{Data acquisition methodology}
% Sizes of all curved folds and grippers were measured by caliper (thickness) and ruler (width and length). For every single printed sample, a mono-domain sample (with linear alignment) was printed at the same time, and later these mono-domain samples were used to either actuate them in toluene (by swelling) or on a hot plate.     
\newpage
\section{Movie captions}

The supplemental movies are available at \url{https://drive.google.com/drive/folders/1CR5TdbZNhveHiDYt0_a20O7_nQYS6xZq?usp=sharing}.

{\bf M1}. Kinematics of a curved-fold origami.

{\bf M2}. Kinematics of a symmetric positive (S+) fold.

{\bf M3}. Kinematics of a symmetric negative (S-) fold.

{\bf M4}.Kinematics of an asymmetric positive (A+) fold.

{\bf M5}. Kinematics of an asymmetric negative (A-) fold.

{\bf M6}. Gaussian gripper: simulation and experiment.

{\bf M7}. A Gaussian gripper lifting a load up to 40x the gripper's own weight.

{\bf M8}. A Gaussian gripper lifting a wide range of objects.

\bibliography{curved_fold}